\input harvmac

\lref\sw{  N.~Seiberg and E.~Witten,
  ``Monopole Condensation, And Confinement In N=2 Supersymmetric Yang-Mills
  Theory,''
  Nucl. Phys. B {\bf 426}, 19 (1994)
  [arXiv:hep-th/9407087];
  N.~Seiberg and E.~Witten,
  ``Monopoles, duality and chiral symmetry breaking in N=2 supersymmetric
  QCD,''
  Nucl.\ Phys.\  B {\bf 431}, 484 (1994)
  [arXiv:hep-th/9408099].
}

\lref\klmvw{ A.~Klemm,
W.~Lerche, P.~Mayr, C.~Vafa and N.~P.~Warner, ``Self-Dual Strings and
N=2 Supersymmetric Field Theory,'' Nucl.\ Phys.\ B {\bf 477}, 746
(1996) [arXiv:hep-th/9604034].
\semi
S.~Katz, P.~Mayr and C.~Vafa,
  ``Mirror symmetry and exact solution of 4D N = 2 gauge theories. I,''
  Adv.\ Theor.\ Math.\ Phys.\  {\bf 1}, 53 (1998)
  [arXiv:hep-th/9706110].
}

\lref\bcov{M.~Bershadsky, S.~Cecotti, H.~Ooguri and C.~Vafa,
  ``Kodaira-Spencer theory of gravity and exact results for quantum string
  amplitudes,''
  Commun.\ Math.\ Phys.\  {\bf 165}, 311 (1994)
  [arXiv:hep-th/9309140].
}
\lref\agnt{ I.~Antoniadis, E.~Gava, K.~S.~Narain and T.~R.~Taylor,
  ``Topological amplitudes in string theory,''
  Nucl.\ Phys.\  B {\bf 413}, 162 (1994)
  [arXiv:hep-th/9307158].
}
\lref\nek{ N.~A.~Nekrasov,
  ``Seiberg-Witten prepotential from instanton counting,''
  arXiv:hep-th/0306211 \semi
N.~Nekrasov and A.~Okounkov,
  ``Seiberg-Witten theory and random partitions,''
  arXiv:hep-th/0306238.
}

\lref\witten{E.~Witten, ``Solutions of
four-dimensional field theories via M-theory,'' Nucl.\ Phys.\ B {\bf
500}, 3 (1997) [arXiv:hep-th/9703166].
}

\lref\gaitto{ D.~Gaiotto,
  ``N=2 dualities,''
  arXiv:0904.2715 [hep-th].
}

\lref\gaiomalda{  D.~Gaiotto and J.~Maldacena,
  ``The gravity duals of N=2 superconformal field theories,''
  arXiv:0904.4466 [hep-th].
}

\lref\agt{  L.~F.~Alday, D.~Gaiotto and Y.~Tachikawa,
  ``Liouville Correlation Functions from Four-dimensional Gauge Theories,''
  arXiv:0906.3219 [hep-th].
}

\lref\other{  
  N.~Drukker, D.~R.~Morrison and T.~Okuda,
  ``Loop operators and S-duality from curves on Riemann surfaces,''
  arXiv:0907.2593 [hep-th].
\semi
D.~Gaiotto,
  ``Asymptotically free N=2 theories and irregular conformal blocks,''
  arXiv:0908.0307 [hep-th]
\semi
A.~Mironov, S.~Mironov, A.~Morozov and A.~Morozov,
  ``CFT exercises for the needs of AGT,''
  arXiv:0908.2064 [hep-th]
\semi
A.~Mironov and A.~Morozov,
  ``The Power of Nekrasov Functions,''
  arXiv:0908.2190 [hep-th]
\semi
A.~Mironov and A.~Morozov,
  ``On AGT relation in the case of U(3),''
  arXiv:0908.2569 [hep-th]
\semi
 L.~F.~Alday, D.~Gaiotto, S.~Gukov, Y.~Tachikawa and H.~Verlinde,
  ``Loop and surface operators in N=2 gauge theory and Liouville modular
  arXiv:0909.0945 [hep-th]
\semi
  N.~Drukker, J.~Gomis, T.~Okuda and J.~Teschner,
  arXiv:0909.1105 [hep-th].
}

\lref\gv{
  R.~Gopakumar and C.~Vafa,
  ``On the gauge theory/geometry correspondence,''
  Adv.\ Theor.\ Math.\ Phys.\  {\bf 3}, 1415 (1999)
  [arXiv:hep-th/9811131].
}

\lref\dv{ R.~Dijkgraaf and C.~Vafa,
  ``Matrix models, topological strings, and supersymmetric gauge theories,''
  Nucl.\ Phys.\  B {\bf 644}, 3 (2002)
  [arXiv:hep-th/0206255] \semi
R.~Dijkgraaf and C.~Vafa,
  ``On geometry and matrix models,''
  Nucl.\ Phys.\  B {\bf 644}, 21 (2002)
  [arXiv:hep-th/0207106] \semi
 R.~Dijkgraaf and C.~Vafa, ``A perturbative window into
  non-perturbative physics,'' arXiv:hep-th/0208048.}

\lref\civ{F.~Cachazo, K.~A.~Intriligator and C.~Vafa,
  Nucl.\ Phys.\  B {\bf 603}, 3 (2001)
  [arXiv:hep-th/0103067].
}

\lref\adkmv{  M.~Aganagic, R.~Dijkgraaf, A.~Klemm, M.~Marino and C.~Vafa,
  ``Topological strings and integrable hierarchies,''
  Commun.\ Math.\ Phys.\  {\bf 261}, 451 (2006)
  [arXiv:hep-th/0312085].
}

\lref\akmv{  M.~Aganagic, A.~Klemm, M.~Marino and C.~Vafa,
  ``The topological vertex,''
  Commun.\ Math.\ Phys.\  {\bf 254}, 425 (2005)
  [arXiv:hep-th/0305132].
}

\lref\amv{  M.~Aganagic, M.~Marino and C.~Vafa,
  ``All loop topological string amplitudes from Chern-Simons theory,''
  Commun.\ Math.\ Phys.\  {\bf 247}, 467 (2004)
  [arXiv:hep-th/0206164].
}

\lref\rus{A.~Marshakov, A.~Mironov and A.~Morozov,
  ``Generalized matrix models as conformal field theories: Discrete case,''
  Phys.\ Lett.\  B {\bf 265}, 99 (1991)
\semi
S.~Kharchev, .~A.~Marshakov, A.~Mironov, A.~Morozov and S.~Pakuliak,
  ``Conformal Matrix Models As An Alternative To Conventional Multimatrix
  Nucl.\ Phys.\  B {\bf 404}, 717 (1993)
  [arXiv:hep-th/9208044]
\semi
A.~Morozov,
  ``Matrix models as integrable systems,''
  arXiv:hep-th/9502091

\semi
I.~K.~Kostov,
  ``Conformal field theory techniques in random matrix models,''
  arXiv:hep-th/9907060.
}

\lref\mehta{
M. L. Mehta, {\it Random matrices}, Pure and Applied Mathematics Series
142 (2004).}

\lref\bet{
B.~Eynard and O.~Marchal,
``Topological expansion of the Bethe ansatz, and non-commutative algebraic
geometry,''
  JHEP {\bf 0903}, 094 (2009)
  [arXiv:0809.3367].
\semi
 P.~Desrosiers,
  ``Duality In Random Matrix Ensembles For All Beta,''
  Nucl.\ Phys.\  B {\bf 817}, 224 (2009)
 \semi
 A. Zabrodin,
``Random matrices and Laplacian growth'', 
arXiv:0907.4929 [math-ph].  
}

\lref\ovd{ H.~Ooguri and C.~Vafa, ``Worldsheet
Derivation of a Large N Duality,'' Nucl.\ Phys.\ B {\bf 641}, 3 (2002)
[arXiv:hep-th/0205297].
}

\lref\witcs{E.~Witten, ``Chern-Simons Gauge Theory As A String
Theory,'' Prog.\ Math.\ {\bf 133}, 637 (1995) [arXiv:hep-th/9207094].
}

\lref\ovwil{H.~Ooguri and C.~Vafa,
  ``Knot invariants and topological strings,''
    Nucl.\ Phys.\  B {\bf 577}, 419 (2000)
  [arXiv:hep-th/9912123].
}

\lref\dhsv{ R.~Dijkgraaf, L.~Hollands, P.~Sulkowski and C.~Vafa,
  ``Supersymmetric Gauge Theories, Intersecting Branes and Free Fermions,''
  JHEP {\bf 0802}, 106 (2008)
  [arXiv:0709.4446 [hep-th]].
}

\lref\dhs{R.~Dijkgraaf, L.~Hollands and P.~Sulkowski,
``Quantum Curves and D-Modules,''
  arXiv:0810.4157 [hep-th].
}

\lref\wyll{
N.~Wyllard, ``$A_{N-1}$ conformal Toda field
theory correlation functions from conformal N=2 SU(N) quiver gauge
theories,'' arXiv:0907.2189 [hep-th].
}.

\lref\dvs{  R.~Dijkgraaf and C.~Vafa,
  ``N = 1 supersymmetry, deconstruction, and bosonic gauge theories,''
  arXiv:hep-th/0302011.}

\lref\universal{
E.~Brezin and A.~Zee,
``Universality of the correlations between eigenvalues of large random
matrices,''
  Nucl.\ Phys.\  B {\bf 402}, 613 (1993).}

\lref\dst{R.~Dijkgraaf, A.~Sinkovics and M.~Temurhan,
  ``Universal correlators from geometry,''
  JHEP {\bf 0411}, 012 (2004)
  [arXiv:hep-th/0406247].
}

\lref\eynard{
B.~Eynard and N.~Orantin,
``Algebraic methods in random matrices and enumerative geometry,''
  arXiv:0811.3531 [math-ph].
}

\lref\dveynard{
R.~Dijkgraaf and C.~Vafa,
  ``Two Dimensional Kodaira-Spencer Theory and Three Dimensional Chern-Simons
  Gravity,''
  arXiv:0711.1932 [hep-th].
}

\input epsf

\def\figin{\epsfcheck\figin}\def\figins{\epsfcheck\figins}
\def\epsfcheck{\ifx\epsfbox\UnDeFiNeD
\message{(NO epsf.tex, FIGURES WILL BE IGNORED)}
\gdef\figin##1{\vskip2in}\gdef\figins##1{\hskip.5in}
\else\message{(FIGURES WILL BE INCLUDED)}%
\gdef\figin##1{##1}\gdef\figins##1{##1}\fi}
\def\DefWarn#1{}
\def\figinsert{\goodbreak\topinsert}
\def\ifig#1#2#3#4{\DefWarn#1\xdef#1{fig.~\the\figno}
\writedef{#1\leftbracket fig.\noexpand~\the\figno}%
\figinsert\figin{\centerline{\epsfxsize=#3mm \epsfbox{#2}}}
\bigskip\medskip\centerline{\vbox{\baselineskip12pt
\advance\hsize by -1truein\noindent\footnotefont{\sl Fig.~\the\figno:}\sl\ #4}}
\bigskip\endinsert\noindent\global\advance\figno by1}

\def\C{{\bf C}}
\def\R{{\bf R}}
\def\Z{{\bf Z}}

\def\b{\beta}

\def\e{\epsilon}

\def\P{{\bf P}}

\def\d{\partial}
\def\dbar{{\overline\partial}}

\def\Tr{{\rm Tr}}

\def\cO{{\cal O}}

\def\cF{{\cal F}}
\def\cN{{\cal N}}

\def\({\bigl(}
\def\){\bigr)}
\def\<{\langle\,}
\def\>{\,\rangle}

\def\]{\right]}
\def\[{\left[}

\Title
{\vbox{
\vskip-20mm
}}
{\vbox{
\centerline{Toda Theories, Matrix Models, Topological Strings, }
\vskip .2in
\centerline{and}
\vskip .2in
\centerline{$N=2$ Gauge Systems}}}
\vskip .3in
\centerline{Robbert Dijkgraaf$^1$ and Cumrun
Vafa$^{2}$}
\vskip .3in

\centerline{$^1$ Institute for Theoretical Physics, 
University of Amsterdam, The Netherlands}

\centerline{$^2$ Jefferson Physical Laboratory, Harvard University,
Cambridge, MA 02138}

\vskip .2in

We consider the topological string partition function, including the
Nekrasov deformation, for type IIB geometries with an $A_{n-1}$
singularity over a Riemann surface.  These models realize the $\cN=2$
$SU(n)$ superconformal gauge systems recently studied by Gaiotto and
collaborators.  Employing large $N$ dualities we show why the
partition function of topological strings in these backgrounds is
captured by the chiral blocks of $A_{n-1}$ Toda systems and derive the
dictionary recently proposed by Alday, Gaiotto and Tachikawa.  For the
case of genus zero Riemann surfaces, we show how these systems can
also be realized by Penner-like matrix models with logarithmic
potentials. The Seiberg-Witten curve can be understood as the spectral
curve of these matrix models which arises holographically at large
$N$. In this context the Nekrasov deformation maps to the
$\beta$-ensemble of generalized matrix models, that in turn maps to
the Toda system with general background charge.  We also point out the
notion of a double holography for this system, when both $n$ and $N$
are large.


\Date{September 2009}

\newsec{Introduction}

The geometry of ${\cal N}=2$ supersymmetric gauge theories in 4
dimensions has played a prominent role in understanding strong
coupling aspects of gauge theories.  In particular the geometry
underlying the vacuum solution of $SU(2)$ gauge theory with matter
discovered by Seiberg and Witten \sw\ and its
various subsequent generalizations have been a very fruitful arena.
The embedding of these gauge theories in string theory has led to
valuable insights into both gauge theory dynamics as well as into
string dualities. Typically one finds a construction in string theory
that realizes the gauge system and then uses string dualities to find
the vacuum solution.

There are two prominent ways this has been done.  The first is in the
form of geometric engineering of gauge theories \klmvw.  Here one
considers a local ADE geometry of a Calabi-Yau 3-fold in type IIA to
engineer the gauge theory and uses the mirror symmetry to find the
solution of it in terms of the mirror Calabi-Yau of type IIB.  For
some special gauge groups and quivers the mirror CY data is captured
by a Riemann surface which gets identified with the Seiberg-Witten
curve.  Moreover, by another duality, this gets mapped to NS 5-branes
of type IIA wrapping the curve and filling the spacetime \klmvw.  The
topological B-model on the mirror Calabi-Yau not only yields the
vacuum geometry and thus the ${\cal N}=2$ prepotential (given by genus
zero partition function of the topological string), but also computes
induced gravitational corrections (coming from genus $g>0$ topological
string amplitudes) of the form \refs{\bcov,\agnt}
$$
\int d^4\theta \, {\cal F}_g(\mu){\cal W}^{2g}.
$$
Here ${\cal W}$ is the self-dual graviphoton chiral field, and ${\cal
F}_g(\mu)$ denotes the genus $g$ amplitude of topological strings,
written as a function of the complex structure moduli $\mu$ (in the
B-model). All these quantities are elegantly combined in terms of the
topological string partition function
\eqn\topstring{
Z(\mu;g_s) = \exp \sum_{g\geq 0} g_s^{2g-2} {\cal F}_g(\mu) .
}

It was shown in \nek\ that these corrections are also directly
computable in the gauge theory setup using equivariant instanton
calculus of a $U(1) \times U(1)$ subgroup of the rotation group
$SO(4)$ acting on Euclidean space-time.  Moreover these gravitational
corrections were generalized to include a one-parameter deformation
given by the anti-self dual twisted graviphoton field. This refined
partition function $Z(\mu;\e_1,\e_2)$ of Nekrasov reduces to the
topological string \topstring\ in the case $\e_2= -\e_1=g_s$.

In another approach \witten, one realizes the gauge system in terms of
D4 branes suspended between NS 5-branes, and one obtains the strong
coupling solution by lifting the geometry to M-theory, where the
relevant geometry emerges in the form of the M5 brane wrapping the
curve.  In fact, a large class of such gauge theories that are
conformal, together with their mass deformations, were constructed and
studied in \witten\ by considering a system of $N$ M5 branes wrapping
a sphere or a torus.

Recently these constructions were generalized by Gaiotto \gaitto\ to
the case of arbitrary Riemann surfaces, and were also studied
holographically in \gaiomalda.  Furthermore a surprising conjecture
was made by by Alday, Gaiotto and Tachikawa in \agt, that for the case
of two M5 branes, {\it i.e}\ a $SU(2)$ gauge theory, the Nekrasov
partition functions are simply given by the chiral blocks of a
Liouville theory on the corresponding curve. More precisely the AGT
conjecture says that:

\item{(1)} the mass deformations are given by Liouville vertex operators
carrying the corresponding momentum;

\item{(2)} the Liouville momenta in the intermediate chanels characterize the
Coulomb branches of the gauge theory; 

\item{(3)} the position of the vertex operators correspond to gauge coupling 
constants; and

\item{(4)} the choice of the background charge of the Liouville CFT is
related to Nekrasov's deformations $\e_1,\e_2$ of the topological
string partition function.

\medskip

By now there is strong evidence \refs{\wyll,\other} that this
correspondence with Liouville for the $SU(2)$ case, and its
generalization to Toda for the $SU(n)$ case is valid.  The main
question is what is the explanation of such a duality.  In this paper
we offer a simple stringy explanation based on large $N$ dualities of
topological strings.

The basic ingredients of our derivation is as follows: We realize the
gauge theories of interest by geometrically engineering them in type
II theories and ask how one can compute topological string amplitudes
for such backgrounds.  It has been known that for toric geometries a
powerful approach for such computations is through the study of large
$N$ dualities for topological strings \gv, where topological string
amplitudes get related to gauge theories on branes.  In particular in
the type IIA setup this relates topological A-model string amplitudes
to Chern-Simons amplitudes.  For example, the partition function of
A-model on the resolved conifold, gets related to Chern-Simons theory
on $S^3$. The B-model analog of these large N dualities where proposed
in \dv, motivated from their embedding in superstrings
\civ. It was shown how the matrix models on the gauge theory
side, which is the effective theory on the B-branes, compute
amplitudes of the closed string side (where the closed string geometry
was captured by the resolvent of the large $N$ limit of the matrix
model).  Moreover the analog of brane probes correspond to free
fermions living on the spectral curve
\adkmv.

The study of D-brane probes on both sides of the duality leads to a
local relation between open string computations, such as knot
invariants, with string amplitudes in the presence of brane probes.
The topological vertex formalism \akmv\ is a theory which captures the
correlations of such D-branes, which can be computed using this
duality.  The closed string amplitudes are the special case of such
amplitudes where no brane is inserted, {\it i.e.} its vacuum
amplitude.

To use this machinery we are naturally led to ask which CY geometries
realize the ${\cal N}=2$ CFT and their deformations that are relevant
for the AGT conjecture.  It turns out that the simplest examples of
such ${\cal N}=2$ CFT's were already constructed in the type IIA setup
in \amv\ as a precursor to the more general construction embodied in
the topological vertex.  This includes for example the case of
$SU(2)$ with 4 hypermultiplet doublets.  The reason these
constructions were simpler was that by a series of transitions one
could get rid of all the 2-cycles in the type IIA geometry and
describe the entire geometry using transitions for geometries with
3-cycles only, with branes wrapped on them.

However the expansion point suitable for \amv\ corresponds to the
large Coulomb branch parameter for the $SU(2)$, whereas the expansion
relevant for the AGT conjecture involves expansion near zero Coulomb
parameters.  Moreover, the answer in \amv\ is directly relevant for
the 5d theories compactified on a cricle.  To obtain the purely 4d
answer we need to take the limit of small radius of the circle.  This
leads us to the study of the B-model topological string for the
relevant geometry, which is more suitable for the 4d limit. It is thus
natural to look for a simpler version of \amv\ relevant for the AGT
conjecture and that is what we focus on in this paper.

There are two alternative derivations we find.  In the first approach
we utilize the well-known connections between matrix model and
topological strings on Calabi-Yau \dv\ on the one hand, and the
relation between matrix model and Toda systems \rus\ on the other
hand, to derive the AGT conjecture.  Namely, we use the matrix model
directly to engineer the corresponding Calabi-Yau. Here the CY of
interest is obtained by geometric transitions from the geometry on
which the branes live, induced by having a large number of branes
(corresponding to the large $N$ limit of matrix model).  This
derivation is simplest for an $A_{n-1}$ singularity over a sphere with
punctures, although it can, with some subtleties, be generalized to
arbitrary curves. 

In this derivation, the starting geometry before transition
corresponds to a subspace of the masses and Coulomb branch parameters
where the Toda momenta are additively conserved.  This Calabi-Yau
geometry is fixed in terms of the data of this restricted masses and
Coulomb parameters. In the case of $SU(2)$ this geometry involves a
number of conifold points.  Blowing the conifold points and placing
topological B-branes in this background give rise to a matrix model,
whose action can be read off from the CY geometry.  The open string
action involves a sum of Penner-like potentials.  The momentum
non-conserving part of the Liouville amplitudes, which lead to more
general mass and Coulomb moduli, arise by going to the large $N$ limit
of the matrix model and distributing the matrix eigenvalues at the
different critical points (corresponding to deformations of the
conifolds).  The correlation functions of the corresponding Toda
theory get mapped to $A_{n-1}$ quiver matrix model. The well known
direct map between matrix models and Liouville theories \rus\
completes the derivation of AGT.

Our second derivation, namely the brane probe approach, is more
conceptual. We work with the brane probes of CY geometries and find a
nice way to study their effective action near the $A_{n-1}$
singularity, where the Coulomb parameter is near zero.  For such
geometries instead of one type of D-brane probe we get $n$ of them.
Moreover, the brane probes, which are well known to relate to
fermions \adkmv, now become a system of $n$ free fermions.
Bosonization of these fermions leads to the $n$ Toda fields of the
$U(n)$ theory.  It is important that we have an extra $U(1)$ in the
theory which resolves some of the puzzles in AGT prescription
associated with a missing $U(1)$.  However, it turns out that the fact
that these $n$ fermions live on the $A_{n-1}$ singularity makes them
interacting in an interesting, non-commutative way.  This involves
deforming the theories by insertion of operators which correponds to
currents of the $U(n)$ theory along cycles.  This interaction gets
mapped, upon bosonization, to the Toda potential.  Moreover
introduction of mass parameters corresponds to a geometry, which
before transition simply multiplies the fermionic wave function by a
pole term, with residue given by mass.  We show how this arises and
explain why it maps, upon bosonization, to the vertex operator for the
Toda field with momentum in the fundamental weight of the corrsponding
gauge group. 

Furthermore we explain that the Nekrasov deformation of topological
string amplitude correspond to the spectral flow operator for the
fermionic sea.  In the matrix model approach, the Nekrasov deformation
gets mapped to considering a `$\beta$-ensemble' \mehta\ (for some
recent literature see \bet) where $\beta=-\epsilon_2/\epsilon_1$ and
the Vandermonde determinant is raised to the powers of $2\beta$ instead
of 2.

The organization of this paper is as follows: In section 2 we recall
the relation between matrix models and topological strings on the one
hand, and matrix models and Toda theories on the other.  Using this we
show how the AGT conjecture arises for genus 0, and can be extended to
higher genera.  In section 3, we start the alternative brane probe
derivation by recalling the relevant Calabi-Yau geometries, both in
the type IIA set up and their mirror in type IIB.  In addition we
recall some basic facts about brane probes for topological strings and
their relation to fermions.  In section 4, we discuss branes for the
$A_{n-1}$ singularities and how bosonization of the system relates it
to Toda.  Furthermore we describe geometries which `chop-off' the
$A_{n-1}$ singularities to pieces and their effect on D-brane probes.
This leads us to the alternative derivation of AGT conjecture.  In
section 5 we describe the Nekrasov deformation, both in the brane
picture (which leads to background charge for Toda theory) and in the
matrix model approach.  Finally in section 6 we end with some issues
for further study.

\newsec{Matrix Model and the Emergence of Geometry}

Matrix models are known to provide powerful descriptions of topological
string B-models on special classes of Calabi-Yau geometries. We will
see that this technique can also be used in the case of the geometries
that are relevant for ${\cal N}=2$ superconformal gauge theories.
Matrix models thus provide a direct bridge between the Calabi-Yau
geometries of interest, and Liouville and Toda systems.

This connection to matrix models is not surprising. On the one hand we
already know, through geometric transitions, that matrix models do
describe local topological B-models \dv.  On the other hand it is also
known that the collective field for matrix models is related to Toda
systems \rus.  The connection between the two links the rank $N$ of
the matrices with the number of insertion of screening operators on
the Toda side. The new ingredient will be that the correlation
functions of the Toda system are naturally considered in the large $N$
limit in order to connect to the gauge theory quantities.

We begin the discussion below with the case of the single matrix model
and recall its connection with local Calabi-Yau geometries, and at the
same time connect it to the Liouville description.  We then discuss
its extension to multi matrix models.  Note that compared to the
applications considered in \dv, where the number of matrix fields was
related to the number of gauge factors in the quiver, in the current
applications the number of matrices gets related to the rank of the
gauge group.  In geometric engineering terminology this is the same as
exchange of the base and the fiber geometry.

\subsec{Matrix models and CFT}

For convenience let us concentrate first on the $SU(2)$ case that
corresponds to a single $N \times N$ matrix. The
generalization to $SU(n)$ is rather straightforward. In general the
partition function of such a matrix model takes the form
\eqn\mm{
Z = \int_{N \times N} \!\!\!\!\! d\Phi\ \exp {1\over g_s}\Tr\, W(\Phi).  
}
Here and subsequently, all variables are complex and the integral is a
suitable contour integral in the space of complex matrices, {\it e.g.}
the real subspace of hermitean matrices.  The matrix model describes
topological B-model string theory on a Calabi-Yau geometry of the form
$$
uv + F(x,z)=0,
$$
where the equation $F(x,z)=0$ describes some algebraic curve $\Sigma$
in $\C^2$. In the matrix model $\Sigma$ is identified with the
spectral curve and is given by a double cover of the $z$-plane of the
form
\eqn\hyperell{
F(x,z) = x^2 - W'(z)^2 + f(z)=0.  
}
Saddle points of the matrix integral correspond to a distribution
$N=N_1+N_2+\ldots$ over the critical points $y_1,y_2,\ldots$ of the
potential $W(z)$. Such a saddle point determines the so-called quantum
correction $f(z)$ in \hyperell. 

This connection between matrix models and CY geometries
was explained in \dv .  The idea was based on geometric transition
where one considers the geometry before transition given by
$$
uv + x^2-W'(z)^2 =0,
$$
and the conifold singularities at $u=v=x=W'(z)=0$ are resolved into
${\bf P}^1$'s.  We then distribute $N$ copies of B-branes over these
${\bf P}^1$'s and this determines the geometry after transition. 

The relation between matrix model and the topological B-model
on \hyperell\ is simply the large $N$ duality in the context of
topological string. The open B-model is given by a matrix model with
potential $W(\Phi)$ and thus the partition function of the matrix
model computes the closed B-model topological string amplitudes on the
emergent geometry given by \hyperell.  In other words, the matrix
model computes the amplitudes of the closed topological string on a CY
which has undergone a geometric transition.

By tuning the potential and the quantum correction, a general
hyperelliptic curve $x^2 = P(z)$ can be engineered. The higher genus
corrections can be captured in terms of collective string field theory
$\phi(z)$ living on the spectral cover. This scalar field gives a
natural connection to two-dimensional chiral conformal field
theory. In fact, it is known that this collective field for matrix
models is closely related to Liouville theory \rus. Let us briefly
review this connection.

The matrix model partition function can be written in terms of the
eigenvalues $z_I$ of the matrix $\Phi$ as
$$
Z = \int d^N\!z \prod_{I < J}\left(z_I
- z_J\right)^2 \exp \sum_I {1\over g_s} W(z_I)
$$
This expression can be recognized as a correlation function of a free
conformal scalar field theory \rus. In particular, the Vandermonde
determinant of the eigenvalues has an interpretation in terms of an
(integrated) N-point function of vertex operators $e^{i\phi}$ at
positions $z_I$, and the potential term looks like a background gauge
field coupling to the current $\d\phi$. So, we can rewrite the matrix
integral as
\eqn\kn{
Z = \langle N | \int d^N\!z \,e^{i\phi(z_1)} \cdots e^{i\phi(z_N)}
\exp \left(\oint_\infty dz\, {1\over g_s} W(z) \d\phi(z) \right)| 0 \rangle
}
Here $| N \rangle$ denotes the vacuum with total momentum/charge $N$,
$$
\oint \d\phi  | N \rangle = N  | N \rangle.
$$
Correlation functions like \kn\ with a total net charge $N$ we will
denotes as $ \langle \cdots \rangle_N$. Using this notation the
partition can be written in exponentiated form as the generating
function
$$
Z = \left\langle \exp \left( \int_{-\infty}^\infty
dz\,e^{i\phi(z)}\right) \exp \left(\oint_\infty dz\, {1\over g_s}
W(z)\del\phi(z) \right)
\right\rangle_{\strut N}
$$

The scalar $\phi(z)$ appears directly in the large $N$ limit of the
matrix model as the collective field of the eigenvalues. Matching the
normalizations, it is given by\foot{Here normalizations differ. The
Liouville field has a kinetic term with coupling one; the collective
field of matrix models has a coupling $1/g_s^2$ appropriate for closed
string field. So, in general: $\phi_{Liouville} =
g_s \phi_{string}$. We sometimes switch between these conventions.}
$$
\phi(z) = {1\over g_s} W(z) + 2\, \Tr \log(z- \Phi).
$$
or, equivalently,
$$
\d\phi(z) = {1\over g_s} W'(z) + 2\, \Tr{1 \over z- \Phi}.
$$

At this point it is perhaps also good to recall for later use that the
loop equations of the matrix model, which give rise to the spectral
curve \hyperell, are a reflection of the invariance of the matrix
integral \mm\ under infinitesimal diffeomorphisms $\delta \Phi
= \Phi^{n+1}$. These transformations are generated by the modes $L_n$
of the stress-tensor $T(z) = {1\over 2}(\d\phi)^2$.
The spectral curve can thus be written as
$$
x^2 = g_s^2 \langle \d\phi(z)^2 \rangle.
$$

It is convenient to rewrite this CFT representation using free
fermions. Hereto we consider two chiral Dirac fermions $\psi_1,\psi_2$
and their conjugates $\psi^*_1, \psi^*_2$. These can be bosonized in
terms of two scalar fields as
$$
\del\phi_1 = \psi^*_1\psi_1, \qquad \del\phi_2 = \psi^*_2\psi_2. 
$$
In fact, it will be convenient to introduce the odd and even
combinations under the $\Z_2$ exchange $\psi_1 \leftrightarrow \psi_2$
$$
\phi_\pm = \phi_1 \pm \phi_2.
$$
The fermions $\psi_i,\psi_i^*$ form a level one representation of the
$U(2)$ affine current algebra. The $U(2) = U(1) \times SU(2)$ currents
acting on the two fermions can also be bosonized. The $U(1)$ current
is given by $\d\phi_+$ and essentially decouples from the problem,
whereas the triplet of $SU(2)$ currents is expressed in terms of the
field $\phi_-$ as
$$
J_+ = e^{i\phi_-},\qquad J_3 = \d\phi_-, \qquad J_- = e^{-i\phi_-}.
$$
We will now concentrate on the $SU(2)$ sector and the field $\phi_-$,
often dropping the $-$ subscript. 

In terms of these $SU(2)$ fermion currents the matrix model takes the
form
$$
Z = \left\langle \exp \left(\int dz \, J_+(z) \right)
\exp\left(\oint_\infty dz\,  W(z) J_3(z)/g_s\right) \right\rangle_{\strut N}
$$
Notice that the two contours along which we integrate the currents are
different. We integrate $J_+$ along the real axis, whereas $J_3$ is
integrated around $\infty$.

Now, general one-matrix model leads to general hyperelliptic curves
like \hyperell, but here we are actually interested in the simple $A_1$
geometry
$$
uv + x^2=0,
$$
with $z$ a free parameter.  So the Calabi-Yau geometry is
$A_1 \times \C$, with $z$ being the local parameter on the line
$\C$. The corresponding curve is in that case simply given by
$$
F(x,z)=x^2=0,
$$
or its deformation
$$
x^2 - p^2 = 0. 
$$

From the point of view of the matrix model this $A_1$ geometry can be
obtained as a limiting case (double scaling limit) of \hyperell\ with
vanishing matrix potential $W=0$. To this end one introduces a small
quadratic potential $W=\e z^2$, fills the critical point with $N$
eigenvalues and then takes the simultaneous limit $N \to \infty$ and
$\e \to 0$. This produces an (almost) constant eigenvalue density. This
limit can be thought of as describing the local situation at a point of a
general spectral curve. This construction is well-known in
mathematical physics to capture the so-called universal behaviour of
random matrix models \refs{\universal,\dst}. Pictorially this can be
thought of by taking the Wigner circle and deforming it to a very long
and narrow ellips, ending up with a double line.

In fact, we can be a bit more general and choose the limiting case
$W=pz$. This choice of potential corresponds to a background value for
the scalar field given by $\phi(z) = pz$, or $\d\phi=p$. So, there is
constant background momentum $p$ flowing through the $z$ line. In this
case there is no quantum correction, and the effective curve is given
by
$$
x^2 - p^2 = (x+p)(x-p)=0.
$$
This is the deformed $A_1$ singularity. In the limit $p \to 0$ we
recover the original singularity.

Coming back to the matrix model and its bosonic or fermionic CFT
reincarnations, we learn that the special case of vanishing $W$
corresponds to the geometry $A_1 \times \C$. Note that in terms of the
bosonic conformal field theory this special case corresponds to a CFT
action
$$
S = \int d^2z \, \d\phi\dbar\phi + \int dz\, e^{i\phi(z)}.
$$

\ifig\screening{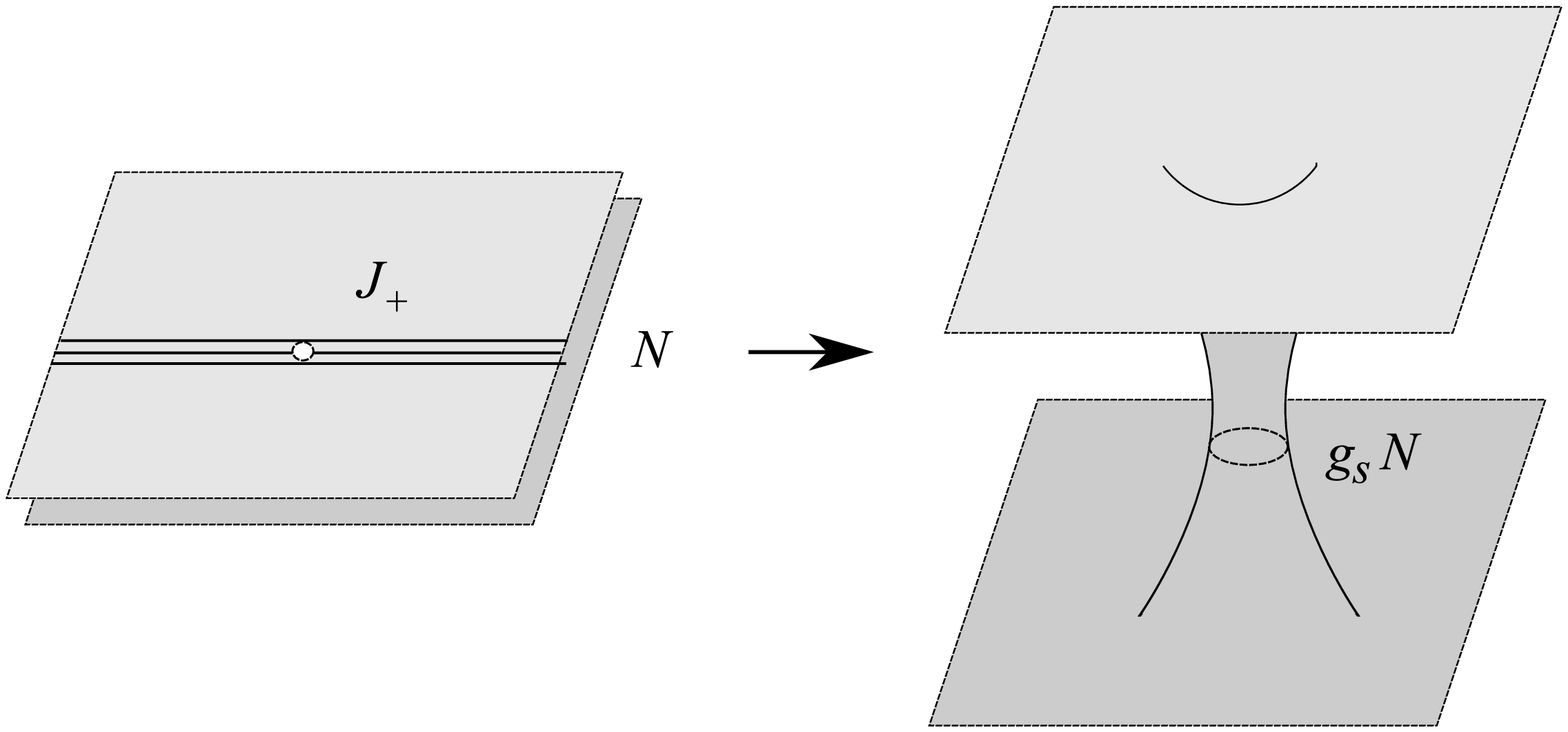}{85}
{Insertions of $N$ copies of the screening charge $\int J_+$ result in
the $N \to \infty$ limit in opening up the double points of the
curve to a size $g_s N$.}

We want to think of this as a chiral version of Liouville theory at
$c=1$ (there is no backgrund charge, yet), where the usual surface
integral of the Liouville potential has been replaced by a line
integral of the screening charge. We think of evaluating this as a
Coulomb gas model, perturbing in the screening charge $\int J_+$,
where the rank $N$ of the matrix model corresponds to the number of
insertions of the screening charge. Note that this sheds a different
light on the spectral curve. The resolution of double points can now
be thought of in terms of condensation of the screening charge
integrals, as illustrated in \screening. In order to make sense of
this picture it helps to consider correlation functions.

\subsec{D-brane insertions and vertex operators}

It is easy to incorporate D-brane insertions in the CFT or matrix
model.  As we will review in section 3, in the fermionic model placing
$m$ B-branes corresponds to the insertion of the operator
$$
V_m  = e^{im\phi_1}=e^{im \phi_+/2} \cdot e^{im\phi_-/2},
$$
which up to the $U(1)$ factor is equivalent to the insertion of
$$
V_m = e^{im \phi_-/2}.
$$
Here we use a normalization of the kinetic term of the fields
$\phi_{\pm}$ so that the vertex operator $V_m$ has
conformal dimension $m^2/4$. In particular the screening operator is
given by $e^{i\phi_-}$ and has dimension 1.

With this normalization the vertex operator $V_m$ is
given in the matrix model by
$$
V_m(q) = \det(\Phi-q)^{m}.
$$
So, we find that a general correlator of D-branes in the topological
B-model of the $A_1 \times \C$ geometry gives rise to the matrix
integral of the form
$$
\langle V_{m_1} \cdots V_{m_k} \rangle_{\strut N} 
= \int_{N \times N} \!\!\!\!\! d\Phi\ \prod_{i=1}^k \det\left(\Phi-
q_i\right)^{m_i}.
$$
This expression can be written in terms of the eigenvalues $z_I$ of
the matrix $\Phi$ as
$$
\langle V_{m_1} \cdots V_{m_k} \rangle_{\strut N} = \int d^N\!z\, 
\prod_{I < J }\left(z_I - z_J\right)^2
\prod_{i, I} \left(z_I - q_i\right)^{m_i} 
$$
Here we recognize the Coulomb gas formulation. Note that there should
be a charge $m_0$ at infinity given by
$$
m_0 = -\sum_{i=1}^k m_i - N.
$$
So, we are actually dealing with a $k+1$ point function on $\P^1$. The
action of $SL(2,\C)$ on the sphere can be used to put these insertions
in general position.

\subsec{The spectral curve}

Now the question is whether we can use the standard matrix model
techniques to solve these expressions. This turns out indeed to be the
case.

The insertions of the determinants $V_m(q)$ can of course be reformulated
as a logarithmic action
$$
W(\Phi) = \sum_{i=1}^N g_s m_i \log \left(\Phi - q_i\right).
$$
Written in this way it is clear we are dealing with a `multi-Penner'
matrix model, with an action given by $k$ logarithms.  

We now want to study the $g_s$ expansion of these expressions. We can
treat this model as any other random matrix model and study its large
$N$ limit, keeping the 't Hooft parameter $g_s N$ finite. We should
also send the matrix model/Liouville parameters $m_i \to \infty$
keeping the {\it gauge theory} masses
$$
\mu_i = g_s m_i
$$
finite. 

General considerations will tell us that in this limit the genus zero
contribution will be captured by an effective algebraic spectral curve
$\Sigma$. In the present $SU(2)$ case it will be given by a double
cover of the sphere. Let us now describe this curve in more detail.

The curve $\Sigma$ is determined by the distribution of the $N$
eigenvalues over the critical point $W'(y_i)=0$. In this case we have
\eqn\crit{
W'(z) =  \sum_{i=1}^k {\mu_i \over z- q_i}.
}
So there $k-1$ critical points and the saddle-point is determined by
the filling fractions
$$
\nu_i = g_s N_i, \qquad i=1,\ldots,k-1.
$$
Of course, the total sum of the $\nu_i$ is fixed to be
$$
\sum_{i=1}^{k-1}\nu_i = g_s N = - \sum_{j=0}^k \mu_j.
$$

It is not difficult to describe the corrsponding spectral curve. First
of all, as in any matrix model, the quantum correction is given by the
saddle point value of the expression
$$
f(z) = \left\langle g_s \sum_{I} {W'(z_I) - W'(z) \over z_I -z
} \right\rangle_N.
$$
(Here $z_I$ are the eigenvalues.) Since $W'(z)$ is a sum of simple
poles at $z=q_i$, one easily verifies that also $f(z)$ takes that form
\eqn\qcor{
f(z) = \sum_{i=1}^k {c_i \over z - q_i}.
}
The coefficients $c_i$ are determined by the filling fractions
$\nu_i$. They satisfy one constraint, that follows directly from the
relation $\sum_I W'(z_I)=0$,
$$
\sum c_i = 0.
$$
This leaves indeed $k-1$ free combinations of the parameters $c_i$
that can be traded for the filling fractions $\nu_i$. 

\ifig\effegeom{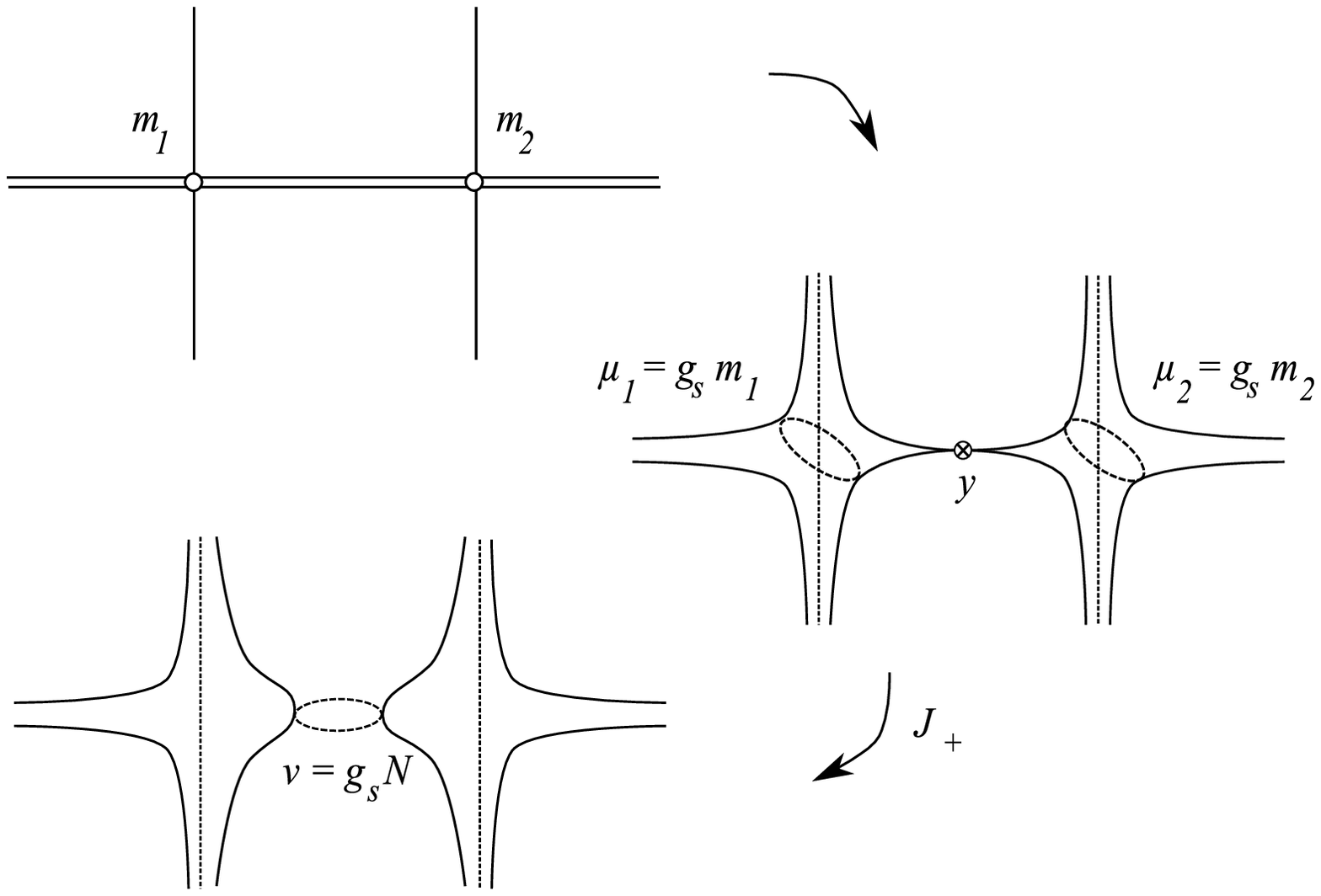}{100}
{Condensation of eigenvalues at the double points of the curve $x^2 =
W'(z)^2$ with $W'(z) = \sum \mu_i/(z-q_i)$ opens them up into branch
points of size $\nu_i = g_sN_i$. In Liouville theory this corresponds
to the inclusion of the screening operator $\int J_+$.}

These parameters have a nice geometric description as depicted in
\effegeom. 
They open up the double points $z=y_i$ of the curve $x^2 = W'(z)^2$
into branch points, creating a higher genus curve. The parameters
$\nu_i$ measure the size of the branch cuts
\eqn\inta{
\oint_{{\cal C}_i} g_S \d\phi = \oint_{{\cal C}_i} xdz = \nu_i,
}
where the contours ${\cal C}_i$ encircles the $i$th cut.

Combining \crit\ and \qcor\ we find that the spectral curve takes the
form
\eqn\cover{
x^2 = {P_{2k-2}(z) \over \Delta_k(z)^2}.
}
with 
$$
\Delta_k(z) =\prod_{i=1}^k (z- q_i).
$$

This gives the right counting of moduli. The numerator $P_{2k-2}(z)$
is a general polynomial of degree $2k-2$. Its $2k-1$ coefficients are
the moduli of the curve. They encode two types of data. First of all
we have the $k+1$ residues $\mu_i^2$ of the double poles of the
quadratic differential $\phi_2(z)$ at $z=q_i$. Here we should also
include the double pole at $z=\infty$ with residue $\mu_0^2$.

Secondly, we have the $k-1$ independent coefficients $c_i$, or
equivalently the moduli $\nu_i$. Since the sum $\sum \nu_i$ is fixed in
term of the masses, this gives $k-2$ extra moduli
$$
a_i = \nu_{i+1}- \nu_i,\qquad i=1,\ldots,k-2.
$$
These $a_i$ have an interpretation of Coulomb parameters of the
corresponding $\cN=2$ gauge theory.

So, let us summarize what we have learned as far
as the Liouville theory is concerned. We considered the
correlation function of vertex operators $e^{im_i\phi}$ in $c=1$
Liouville theory.  The insertions of the Liouville potential $\int
e^{i\phi}$ can be studied in the large $N$ limit, where $N$ is the
number of insertions, if we take at the same time the $m_i$
large. These insertions cluster around the critical points of the
potential induced by the vertex operators, pinning the
eigenvalues. These clusters lead to branch cuts opening up, leaving us
with an {\it effective} CFT living on the spectral cover \cover.\foot{ By
general arguments the scalar field living on the spectral curve is the
Kodaira-Spencer field studied by Eynard and collaborators, see {\it
e.g.} \refs{\eynard, \dveynard}.}

Because of these branch points the scalar field $\phi$ is not
well-defined on the $z$-plane. It transforms as $\phi \to -\phi$ if we
go through the cut. Only on the double cover \cover\ does it become
single valued. In fact, the deck transformation that interchanges the
two sheets acts as $\phi \to - \phi$. This is obvious as $\phi$
started its life as the odd combination $\phi_1 - \phi_2$. The $\Z_2$
involution is simply the action of the Weyl group of $SU(2)$. This is
consistent with the fact that Liouville theory does make sense
downstairs, because here we are daling with the holomorphic square
root or {\it holomorphic blocks}. These typically have monodromies.

Viewed from this perspective the gauge parameters $a_i$ defined
in \inta\ measure the charge flowing through the branch cut or neck
connecting the two sheets. The mass parameters $m_i$ in a similar
fashion measure the charge flowing through the long thin neck that
corrresponds to a puncture.

\subsec{Higher genus curves}

A similar story appears for higher genus curves.  Again we restrict
our attention to the $A_1$ case.

The description can be stated as follows. The Calabi-Yau geometry is
given by a family of $A_1$ singularities fibered over a Riemann
surface $\Sigma$ with local coordinate $z$.   So, to begin with, we
have a geometry of the form
$$uv+x^2=0,$$
for arbitrary $z$ on the curve.  This is dual to two M5 branes wrapping
$\Sigma$.   We next consider deforming the CY geometry to
$$
uv + x^2 - W'(z)^2=0.
$$
where $W'(z)dz$ is a well defined one-form on $\Sigma$.  In particular
we choose $k$ points on $\Sigma$ where $W'(z)$ has a pole with residue
$m_i$.  This corresponds to the bifundamental masses and also `partitioning' the curve
to gauge factors.   Note that the curve is so far factorizable:
$$
x^2-W'(z)^2=0.
$$

To obtain the most complete type IIB geometry for this gauge system we
should also allow arbitrary Coulomb parameters.  We split this into two
parts: The Coulomb branch parameters which keep the curve factorized,
and the ones that do not.  The Coulomb branch parameters
can be realized by having loop momenta flowing through each of the $g$
A-cycles of the Riemann surface.  Let $\omega_j$ denote a basis for
the $g$ holomorphic 1-forms; $\omega_j$ has period 1 around the cycle
$A_j$ and 0 around the other cycles $A_k$.  Furthermore, let the
momentum flowing through the cycles $A_j$ be $p_j$. Including both the
masses and the momenta through A-cycles we find that
$$
W'(z) dz = \sum_{i=1}^k m_i \d_z \log E(z,q_i) + \sum_{j=1}^g p_j \omega_j.
$$
Here $E(z,q_i)$ is a `function' on the Riemann surface which vanishes
to first order as $z\rightarrow q_i$ and is known as the prime form.
This is the effective geometry as seen by the free part of the
Liouville field.  We are now left to deform these further to the most
general Coulomb parameters.  Instead, we {\it induce} these
deformations by large $N$ transition, as in genus 0 case.  Namely we
note that our CY has conifold singularities and resolving the
singularities and placing branes there gives an effective matrix model
as the open string sector of the theory.  The choice of distributing
the eigenvalues of the matrix at different critical points of 
$W'(z)$ gives us the extra Coulomb parameters.  However now we use the
fact that matrix model partition function (the open string side) gives
the closed topological string partition function, {\it i.e.}, the partitions
function for CY with the full general Coulomb parameters, which is
also dual to 5-branes on the corresponding curve.  Finally, the
equivalence of this generalized matrix model with Liouville, leads us
to the derivation of the AGT conjecture (modulo the Nekrasov
deformation which will be covered in section 5).

Note that as in \dv\ placing branes at the resolved $S^2$ gives a
corresponding potential $W(z)$ as input for a generalized matrix
model.  These are generalized matrix models in the sense that the
eigenvalues of the matrix live on the curve $\Sigma$.\foot{In
principle such matrix models can be described by going to a cover
theory, such as the upper half plane, and modding by suitable
subgroups needed to construct $\Sigma$.  This makes it clear why the
eigenvalues, which describe normal deformations of the B-brane, live on
the curve.}  Note that the Vandermonde determinant gets replaced by
the corresponding Green's functions on $\Sigma$ that compute the
$N$-point functions of $J_+$ insertions of the Liouville theory. As
far as the perturbative expansion is concerned it doesn't matter much
how we pick the contours of the $J_+$ integrations, as longs as they
go through all critical points.

This gives the right number of moduli. Note that $W'$ has $k+2g-2$
zeroes as it is a one-form on a genus $g$ curve with $k$ poles. Each
such zero gives rise to a conifold geometry.  However we are free to
choose arbitrary insertions of $\int J_+$ over contours that go
through critical points of the $W'$.  There are $k+2g-3$ independent
cycles. (The `net' number of branes is zero, and thus we have one less
modulus than the number of critical points of $W$.)  In this way, as
we noted before, we end up making the conifold geometry undergo a
transition and we open up such cuts.  This gives altogether $k+2g-3$
additional parameters, depending on the filling fractions.  Together with the original $g$ momenta
$p_i$ we end up with a total of $k+3g-3$ parameters, which is the
expected number of parameters for the Coulomb branches of the gauge
system.

\subsec{Multi-matrix models and Toda systems}

It is not difficult to generalize this picture from $SU(2)$ to
$SU(n)$, in which case we are dealing with an $A_{n-1}$ singularity
$$
uv + x^n =0.
$$
The corresponding Riemann surface is given by
$$
F(x,z)=x^n=0,
$$
and its general deformation is
$$
\prod_{a=1}^n \left(x - p_a\right) =0.
$$

Spectral curves that are $n$-fold covers naturally emerge from quiver
matrix models. In fact, these models can be constructed for any ADE
type singularity. Let us briefly summarize this construction. If the
Lie algebra has rank $r$, we have $r$ matrices $\Phi_a$ of rank $N_a$,
one for each node of the Dynkin diagram, with bifundamental fields
connecting them. Each matrix $\Phi_a$ has an individual potential
$W_a$. After integrating out the bifundamentals, the partition
function can be written in terms of the corresponding eigenvalues
$z_{a,I}$ as (here $a=1,\ldots,r$ and $I=1,\ldots, N_a$)
\eqn\quiver{
Z = \int \prod_{a,I} dz_{a,I} \!\!\! \prod_{(a,I) \not =
(b,J)} \!\!\! \Bigl(z_{a,I}- z_{b,J}\Bigr)^{e_a \cdot e_b} 
\exp \sum_{a,I} {1\over g_s} W_a(z_{a,I}).
}
Here $e_a$ are the simple roots and $C_{ab}=e_a \cdot e_b$ is the
Cartan matrix.

In the case of $A_{n-1}$ quiver matrix model there are $r=n-1$
matrices and the CY geometry takes the form \ref\CachazoGH{
  F.~Cachazo, S.~Katz and C.~Vafa,
  ``Geometric transitions and N = 1 quiver theories,''
  arXiv:hep-th/0108120.
}
$$
uv + \prod_{a=1}^{n} \bigl(x- t_a(z)\bigr) + \ldots = 0.
$$
Here $t_1=0$ and $t_a = \sum_{b=1}^{a-1} W_a'(z)$, and the ellipses
denote the quantum resolution of the double points. Again, taking the
limiting case of linear potentials gives rise to the geometry of the
deformed $A_{n-1}$ singularity
$$
uv + \prod_{a=1}^n \left(x - p_a\right) = 0.
$$

This multi-matrix integral can be similarly cast in a CFT notation,
where now we have $r$ free chiral scalar fields. It is convenient to
use a basis $\phi_a(z)$ that corresponds to the simple roots of the
Lie algebra. We then recognize that the integrand in
expression \quiver\ has the form of a correlation function of a gas of
vertex operators $e^{i\phi_a}$ at positions $z_{a,I}$.  
\eqn\kn{
Z = \left\langle 
\int \prod_{a,I} dz_{a,I} \prod_{a,I} \exp {i\phi_a(z_{I,a})}
\right\rangle_{\strut \{ N_a\}}.
}
Here $N_a$ denotes the total charge, which is now also an
$r$-dimensional vector. The generating function for these correlators
is given by the $c=n-1$ Toda potential
$$
\left\langle \exp \int dz \sum_a e^{i\phi_a(z)} \right\rangle_{\strut \{ N_a\}}.
$$

Of course, for the $A_{n-1}$ case there is a similar fermionic
description in terms of $n$ Dirac fermions $\psi_k,\psi^*_k$. This
gives a realization of level one $U(n)$ Kac-Moody algebra. In the
geometric set-up these fermions can be seen to live on the $n$ sheets
described by the deformed $A_{n-1}$ singularity, as we will discuss in
detail in the next section.

In a similar fashion as for the $SU(2)$ case vertex operator
insertions can be included.  Chopping off the $A_{n-1}$ singularities,
together with accompanying mass insertions, is equivalent to the
insertion of vertex operators of the Toda field with momentum in the
fundamental weight equal to
these masses.  In this way we relate to Toda correlation functions. They
can be computed in the large $N$ limit following the methods described
above. Resolving the $A_{n-1}$ singularities is captured by momentum
flow in the intermediate channel and is obtained by bringing down
appropriate number of Toda potential insertions from the action. These
insertions in turn opens up cuts leading to conifold like transitions
in the geometric setup.

\subsec{Some simple examples}

As we sketched, the relations of gauge theories, Toda systems and
generalized matrix models is universal and can be applied to general
curves. Here we will illustrate this construction with some simple
examples of three and four vertex insertions on the
sphere. Although we haven't shown yet how the background charge of the
Toda field can be generated --- we will return to this in section 5
--- this will not change the geometric picture.

Let us first turn to the three point function. In the gauge theory
this corresponds to the $T_2$ geometry (pairs of pant). In this case
we have two insertions in the $z$-plane, say of mass $m_1$ at $z=0$ and
of mass $m_2$ at $z=1$. This implies a mass at $z=\infty$ of $m_0$,
with $m_0 + m_1 + m_2 + N=0$. In the Liouville model we are computing
the correlation function
$$
\bigl\langle V_{m_0}(\infty) V_{m_1}(0) V_{m_2}(1) \bigr\rangle.
$$
in the scaling limit $m_i = \mu_i/g_s$, $ g_s \to 0$.

The potential is a double Penner model,
$$
W(z) = \mu_1 \log \Phi + \mu_2 \log (\Phi -1).
$$
There is a single critical point $z=y$ solving
$$
W'(z) = {\mu_1 \over z} + {\mu_2 \over z-1}=0.
$$
\ifig\threept{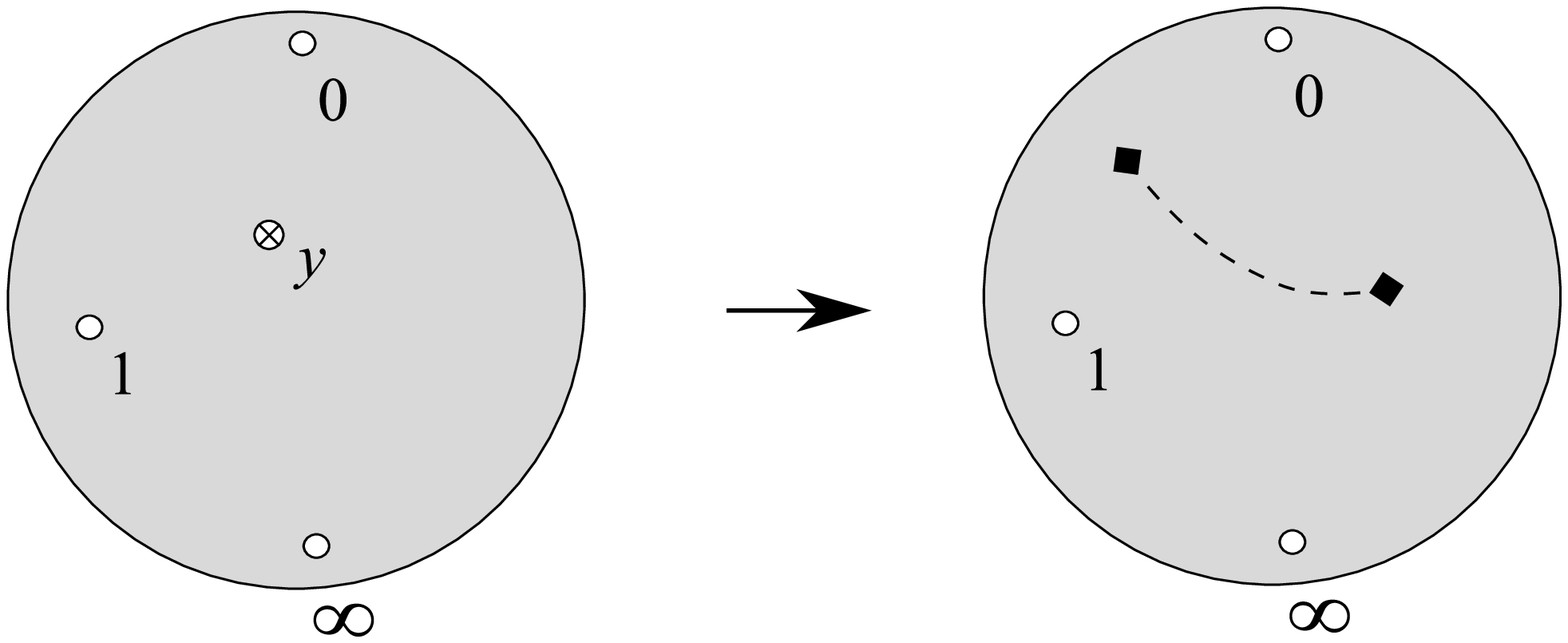}{80}
{The effective geometry of the three point function in Liouville is a
double cover of the sphere with a single branch cut.}
As illustrated in \threept, by condensation of the eigenvalues of the
matrix model or the screening charge insertions, the double point
$z=y$ of the curve $x^2=W'(z)^2$ will be blown up to the form
$$
x^2  = {Az^2 + Bz + C \over \bigl( z(z-1) \bigr)^2 }
$$
Here the coefficients $A,B,C$ are determined by the three masses
$\mu_0,\mu_1,\mu_2$. They also determine the size of the single cut by
$$
\nu = \sum_{i=0}^2 \mu_i.
$$
So we see there are two branch points and the proper place for the
chiral Liouville field $\phi(z)$ is the double cover of the
sphere. Since the cover still has genus zero, this introduces no
Coulomb parameters.

We can similarly consider the four point function.  Here we have three
logarithms
$$
W(z) = \mu_1 \log \Phi + \mu_2 \log (\Phi -1) + \mu_3 \log (\Phi-q).
$$
Of course there is also a mass $\mu_0$ at infinity. The position $q$
gives the UV gauge coupling of the gauge theory. 
\ifig\fourpt{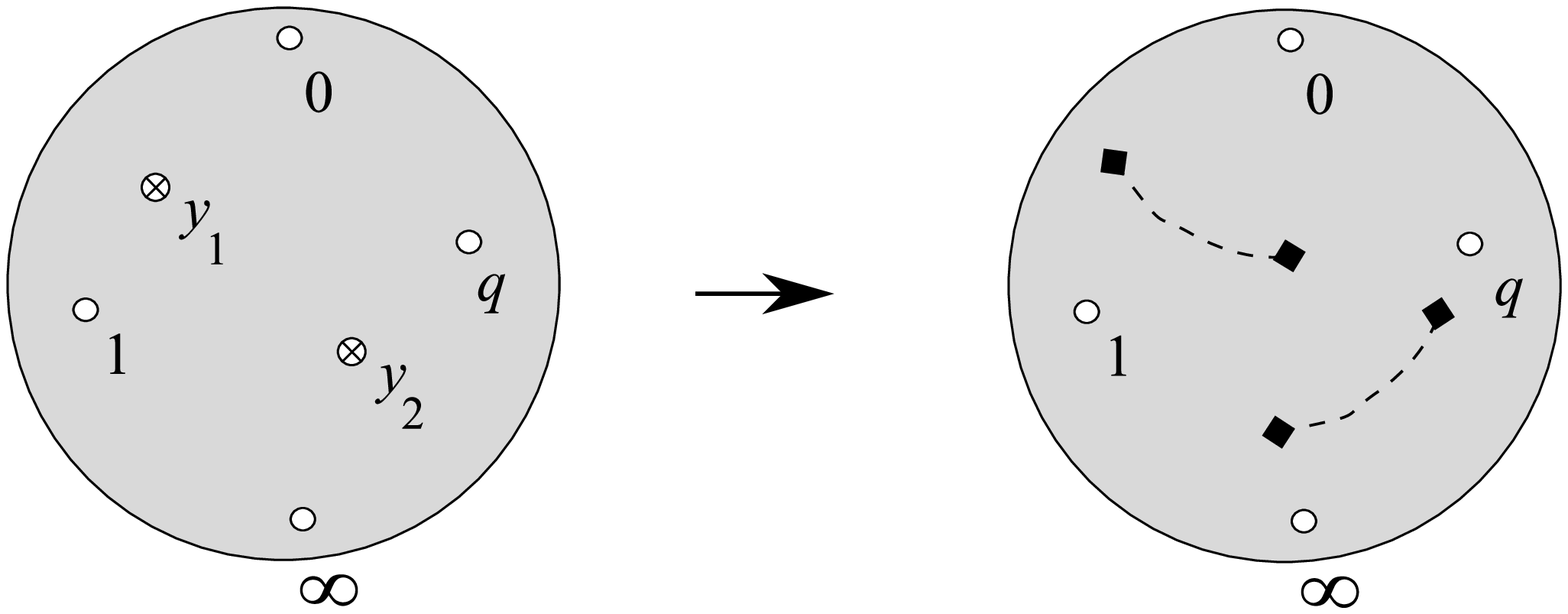}{80}
{The four point function gives a spectral curve with two branch
cuts. Now the cover is an elliptic curve.}

There are now two critical points $y_1,y_2$ that in the $1/N$
expansion each open up to a branch cut of size $\nu_1,\nu_2$, as
illustrated in \fourpt. Here
$$
\nu_1 + \nu_2 = \sum_{j=0}^3 \mu_j.
$$
The double cover
$$
x^2  = {P_4(z) \over \bigl(z(z-1)(z-q)\bigr)^2 }
$$
now has genus one. The 5 parameters in $P_4$ correspond to the 4
masses $\mu_i$ together with the Coulomb parameter $a$, which is given
by the period integral
$$
a= \nu_2-\nu_1 = \oint_{\cal C} xdz
$$
around a contour $\cal C$ that encircles the two cuts in the form of
a figure 8.

\newsec{Local Calabi-Yau Geometries and Large $N$ Transitions}

Beginning in this section we move on to the more conceptual
derivation of AGT conjecture based on B-brane probes.
In this section we review the relevant local CY geometries and 
geometric transitions induced by branes for topological strings. We
will turn to the special case of an $A_{n-1}$ singularity in the next
section.

\subsec{Conifold transitions}

To set the stage let us start with the simplest case of a local
transition which is relevant for us: the conifold transition for a
Calabi-Yau threefold. We consider both the A-model and B-model
versions.  

The singular conifold is given by
$$
uv + xz=0.
$$
In type IIA, one considers the deformed conifold given by
$$
uv + xz = \mu.
$$
\ifig\resconi{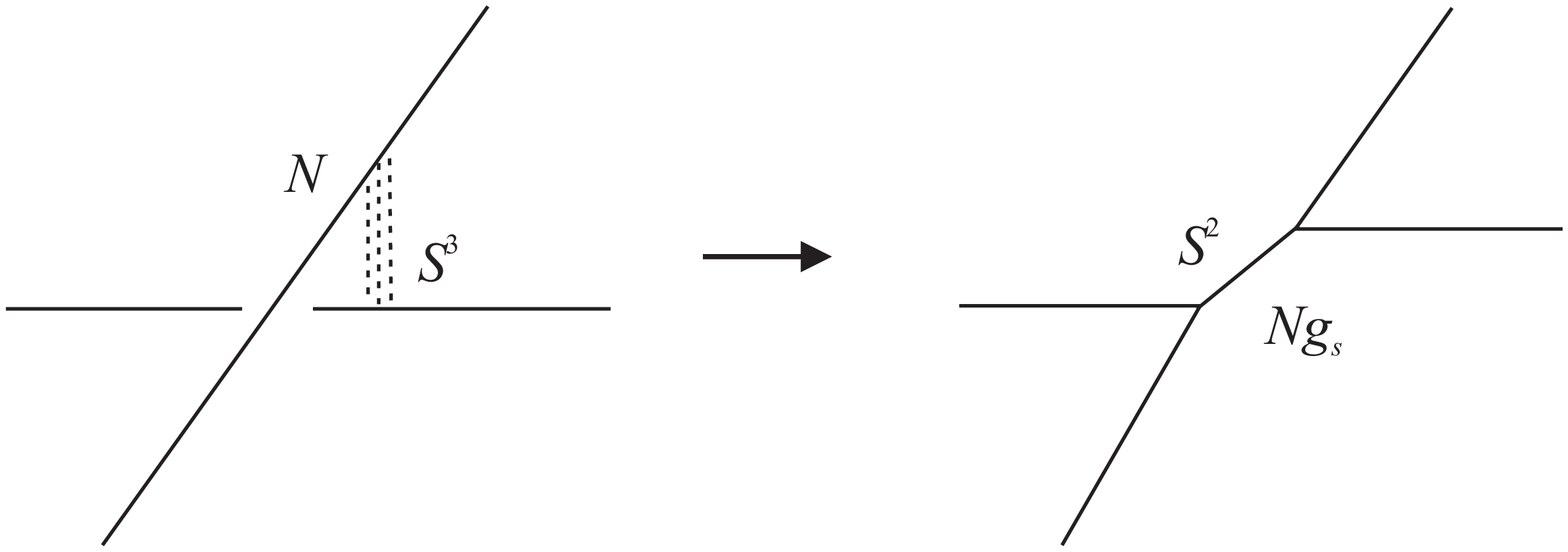}{100}
{The geometric transition in the A-model that resolves the conifold:
 an $S^3$ with $N$ branes shrinks and an $S^2$ of size $g_sN$
 appears.}
This local CY geometry has topology $T^*S^3$. One then wraps $N$
Lagrangian A-branes over $S^3$.  At large $N$ this geometry is
equivalent to topological strings on the resolved conifold, where the
$S^3$ has shrunk and instead an $S^2$ has blown up \gv, see \resconi.
The Kahler class of this $S^2$ is given by $Ng_s$, where $g_s$ is the
topological string coupling constant. The explanation of this
phenomenon is that the Lagrangian $N$ brane produce $N$ units of
`flux' through the $S^2$ surrounding the $S^3$. This flux is measured
by the Kahler form.  A first principle derivation of this duality has
been proposed in \ovd.

The open string degrees of freedom on the $N$ Lagrangian A-branes
gives rise to a $U(N)$ Chern-Simons theory on $S^3$ \witcs.  Thus the
large N duality predicts that the partition function of closed
topological A-model on resolved conifold is equal to the Chern-Simons
partition function on $S^3$.  This statement has been checked in great
detail. Moreover, the duality also works at the level of
observables \ovwil.

In this context the observable on the open string side are Lagrangian
brane probes, intersecting $S^3$ over links, leading to Wilson loop
observables in the CS theory, which on the gravity side are given by
D-brane probes of the resolved conifold geometry.

\ifig\defconi{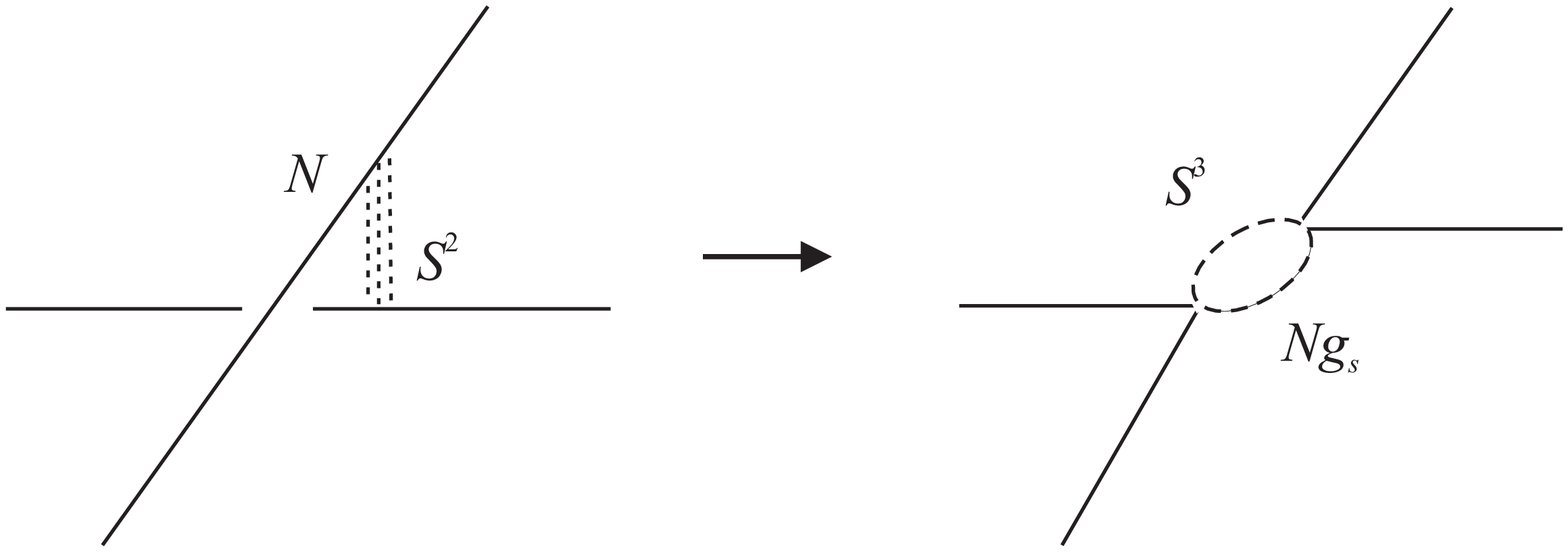}{100}
{In the B-model version the conifold gets deformed. Now an $S^2$
shrinks and an $S^3$ appears.}

The same transition can also be described in the B-model setup, which
is of more interest for the present paper.  Namely in that context one
considers $N$ B-branes wrapped over the $S^2$ of the resolved conifold
geometry, given by the bundle $\cO(-1) \oplus \cO(-1)$ over the
Riemann sphere.  In the large $N$ limit the $S^2$ is replaced by
the $S^3$ of the deformed conifold geometry, where the integral of the
holomorphic 3-form over $S^3$ is given by $Ng_s$ (see \defconi). The
open string side is now given by a matrix model, as we reviewed in the
last section, and the closed string side is given by the B-model
Kodaira-Spencer gravity on the deformed conifold. When reduced to the
underlying curve the dynamical field is the scalar $\phi(z)$.

One can also consider more interesting transitions as were studied in
particular in \amv .  In such cases the geometry on the A-model side
was simply a collection of deformed conifold geometries without any
2-cycles.  Thus one obtains a Chern-Simons gauge theory with a product
group. There is an interaction between the gauge factors given by
annulus diagrams, which lead to insertion of Wilson loop operators in
the product CS theory.

\ifig\fourdoublets{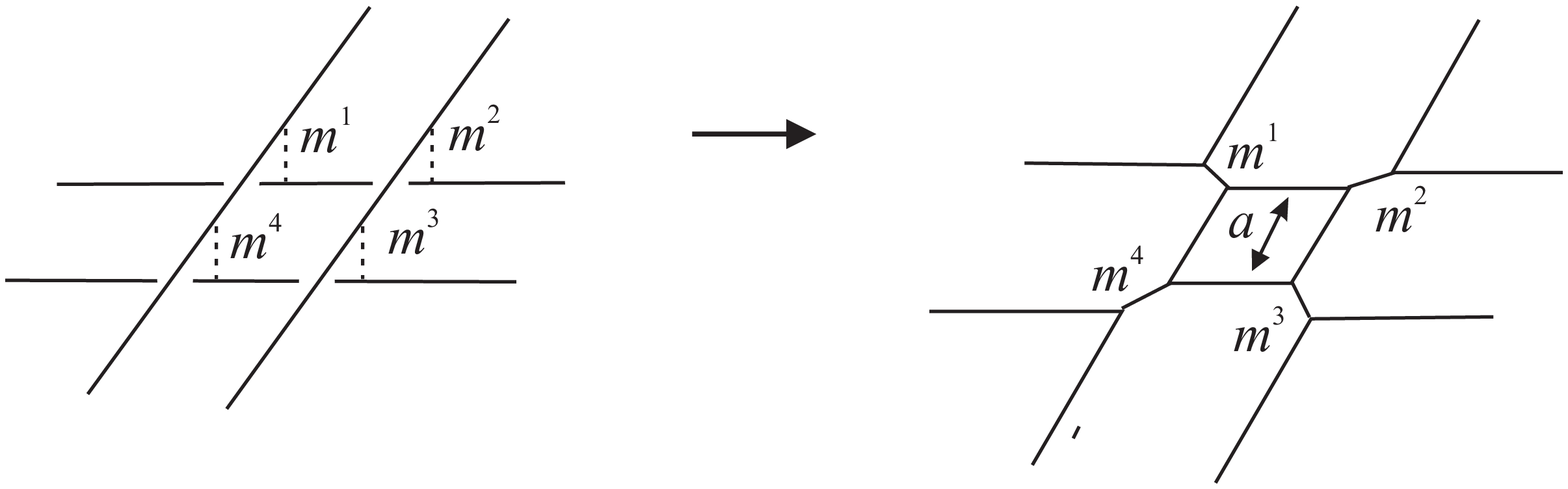}{100}
{Geometry that gives rise to a $SU(2)$ gauge theory with four doublets
with masses $m_1,\ldots, m_4$ and Coulomb parameter $a$.}

This class of theories is particularly useful for finding the large
$N$ dual for theories which in 4 dimensions lead to conformal quivers
of A-type. For example, the geometry depicted in \fourdoublets\ gives
rise to an $SU(2)$ gauge theory with four doublets. To see this note
that the CY geometry after transition is an $ A_1$ singularity fibered
over a $\P^1$ with four blow-ups.  By geometric engineering, each
blow-up corresponds to a hypermultiplet in the fundamental of $SU(2)$,
whose mass $m_i$ is given by the blow up parameter.  One can more
generally consider a product of $k$ factors of $SU(n)$'s with
bifundamental matter fields between them, except at the two ends which
we obtain an additonal $n$ hypermultiplet on each end, in the
fundamental of each of the two end $SU(n)$'s.  These geometries after
transition correspond to `chopping' the $A_{n-1}$ geometries by $k+1$
lines.

\ifig\mirrordoublets{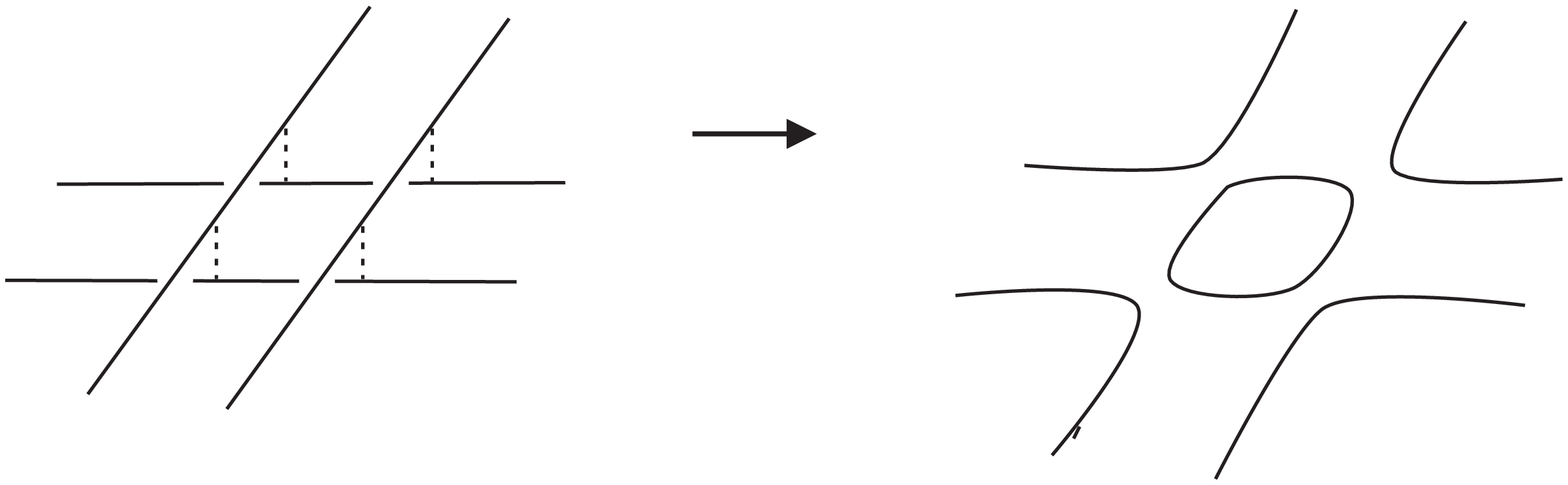}{100}
{The B-model version of the transition of \fourdoublets.}

Similarly we can consider the B-model version of these, which give rise
to the transition depicted in \mirrordoublets.

\subsec{D-brane probes of CY geometries}

D-branes as probes of CY geometries have played a key role in
understanding the dynamics of topological strings.  In the type IIA
setup these brane probes correspond to picking a non-compact
Lagrangian cycle inside the Calabi-Yau and wrap a D6 or D4 brane on
it, where the extra 4/2 dimensions of the brane fill a 4/2-dimensional
subspace of spacetime \ovwil.  These brane probes were useful in
describing Wilson loop observables for the Chern-Simons theory.  A
special class of such Lagrangian branes and their effective action is
what gave rise to the topological vertex \akmv, which in turn led to a
solution of topological A-model for toric 3-folds.  The mirror of
these geometries would correspond to the B-model where D5/D3 branes
wrap a complex curve in the Calabi-Yau and fill a 4/2-dimension
subspace of spacetime.  Since we will be mainly interested in the IIB
setup we will specialize to that case.

As discussed in the previous section, in the type IIB setup the local
geometries of interest are of the form
$$
uv + F(x,z)=0.
$$
The interesting D-brane probes of this geometry correspond to choosing
the following complex 1-dimensional subspace:
$$
u=0,\  v={\rm arbitrary},\ F(x_0,z_0)=0.
$$
The internal geometry of brane probe is one complex dimensional (given
by varying $v$) and its moduli is given by a point $(x_0,z_0)$ on the
$F(x,z)=0$ curve.  Consider the local coordinate on the curve and
denote it by $z$ (where $x$ is solved in terms of $z$ using
$F(x,z)=0$).  

It has been shown that these branes are described by a {\it free}
fermionic system \adkmv. The fermion field $\psi(z)$ descirbes the
creation of a D-brane stretching in the $v$-direction. Similarly the
conjugate field $\psi^*(z)$ describes the creation of a D-brane in the
$u$-direction. Moreover, the Riemann surface is `quantum' in the sense
that on the $x$-$z$ plane there is a
non-commutativity \refs{\adkmv,\dhsv,\dhs} given by
$$
[x,z]=g_s,
$$
where $g_s$ is the topological string coupling constant.  This implies
that for the branes, whose position is given locally by $z$, the
variable $x$ is not quite a classical value, but it is an operator
given by $g_s\cdot \d/\d z$.  Note that this implies that $x$ can be
identified with the Hamiltonian in the {\it chiral} path-integral
where we view $z$ as the chiral `time'.  In other words the chiral
path-integral would be given by
\eqn\ham{{1\over {\hbar}}\int Hdt \rightarrow {1\over g_s}\int x dz}

Viewing the operation by $x$ as the generator of $z$ translations, has
implications for correlation functions of the branes.  In particular
using this, one concludes that the equation for the curve becomes an
operator acting on the partition function of the brane:
$$
F(x=g_s {\d\over \d z} ,z) \cdot 
\Bigl\langle \cdots\psi(z)\cdots\Bigr\rangle=0,
$$
where this is defined up to normal ordering ambiguities in $F$.
Moreover, D-brane probes shift the value of the holomorphic 3-form
$\Omega$ by $Ng_s$, where $N$ is the number of D-branes.  In other
words, $\Omega$ measures the `flux' of the internal 1-complex
dimensional branes (which can be surrounded by a 3-cycle). If we
bosonize the fermions as $\psi = e^{i\phi}$ , then $\Omega $ is
represented by the one-form $\partial \phi$ on the curve.

Since $x$ acts on the brane probe $\psi(z)$ as
$$x\cdot \psi(z)=g_s {\d\psi \over \d z},$$
bosonization implies that
$$
x\cdot \psi(z)=g_s (\partial \phi)\psi(z).
$$
This in particular implies that if we consider a sector where $\phi$
has a momentum $p/g_s$, then $\psi$ is an eigenstate of $x$ with
eigenvalue $p$.

Note that if a 
brane probe goes through a 1-cycle around which we are measuring the
period of $\del \phi$, it can change the period, if its path
intersects it.  Consider in particular the deformed conifold geometry,
represented as the curve
$$F(x,z)=xz-m=0$$
By a change of variables we can view this as
$$x^2=z^2-m$$
This means that we can view the $x$ plane as a double
cover of the $z$ plane, with a cut running between
$z=\pm {\sqrt m}$.  The conifold size is measured by
$$\oint_{cut} xdz =m$$
(up to normalization factors which we ignore here).  Now consider
moving $k$ B-branes from infinity and through the cut to the other
sheet and taking them to infinity of the other sheet.  Then after this
operation we have
$$
\oint_{cut}xdz=m+kg_s.
$$

\newsec{$A_{n-1}$ Geometries and D-brane Probes}

Following these general remarks we will now turn to the geometries that lead
to conformal gauge theories of the type considered in section 3.

\subsec{The non-commutative geometry of $A_{n-1}$ singularities}

We are interested in CY backgrounds described by a family of $A_{n-1}$
singularities fibered over a Riemann surface $\Sigma$. These
geometries are T-dual to a 5-brane wrapping $\Sigma$ and extending in
space-time. They corrspond to supersymmetric gauge theories in four
dimensions with $SU(n)$ gauge group. Locally, the geometry will be
given by
$$
uv =x^n,
$$
where this singularity has a locus given by $\Sigma$, which we
parametrize by a coordinate $z$.  Note that the holomorphic
3-form is given by 
$$
\Omega ={d u \over u} \wedge dx \wedge dz.
$$
In particular $dx$ transforms as a section of $K_\Sigma^{-1}$, the
inverse of the canonical bundle on the surface, so that the two-form
$dx \wedge dz$ has a section over $\Sigma$.

To describe a brane probe in this geometry we choose a point on the
space parametrized by $(x,z)$ that satisfies $x^n=0$.  In other words,
we choose any point on the surface $\Sigma$ and in addition choose a
module for the variable $x$, such that it satisfies the relation $x^n=0$.
Such a module is naturally labelled by a state $\psi_1$ which is the
`lowest' state with $x$-charge.  In other words, $\psi_1$ generates
the rest of the module by the action of $x$ on it (the module is
cyclic)
$$
\psi_1, x\psi_1,...,x^{n-1}\psi_1.
$$
So at each point $z \in \Sigma$ we have an $n$-dimensional module
given by the basis
$$
\psi_i(z) = x^{i-1}\psi_1(z), \qquad i=1,\ldots,n.
$$

These fermions $\psi_i(z)$ represent the $n$ choices of branes. In
analogy with minimal LG theories these branes can be identified with
the vacua of RR sector and $x$ is the chiral field acting on them.  In
this sense we can think of $\psi_1$ as the state with the lowest
R-charge. At a point $z$ the action of $x$ is given by the $n \times
n$ matrix
$$
x = 
\left( \matrix{ 
0 & 0 & \ldots & & 0 \cr
1 & 0 & \ddots & & \vdots \cr
0 & \ddots & \ddots & & \cr
\vdots & \ddots &  & & 0 \cr
0  & \ldots & 0  & 1 & 0 \cr}  \right)
$$
Note that in terms of its action on the fermion states at position
$z$, the operator $x$ is represented by the current
$$
J_+(z) =\sum_{i=1}^{n-1}J_+^i= \sum_{i=1}^{n-1}\psi^*_i\psi_{i+1},
$$
because
$$
\oint_{z} dw J_+(w) \cdot \psi_i(z)=\psi_{i+1}(z).
$$

However, this cannot be the full story. Taking quantum corrections of
the topological string into account, we know that $x$ gets an extra
contribution $g_s \d/\d z$.  This seems to be in contradiction with
the classical statement that $x$ maps one brane to another as
$$
x: \ \psi_i(z) \to \psi_{i+1}(z).
$$
Combining these two ingredients, we see that in the quantum theory $x$
should be represented by a covariant derivative (`Drinfeld-Sokolov
connection')
$$
x = g_s D_z,
\qquad
D_z =  \d_z + {1 \over g_s} J_+(z).
$$
Given the relation \ham\ this implies that the chiral path-integral, due to
this covariantization, picks up an additional term given by the insertion of
$$
\exp \int dz \, J_+(z).
$$
This can be seen as a background gauge field 
that restores the action of $x$ on the path integral.  In other words,
now there is an `internal' action of the field $x$ on the branes
captured by the non-trivial connection $J_+$. It is the presence of
$J_+$, and mainly this, that reflects that we are dealing with an
$A_{n-1}$ singularity. Note that the screening operator is naturally
mapped to the chiral part of the $A_{n-1}$ Toda potential:
$$
J_+ = \sum_{i=1}^{n-1}\psi_i^*\psi_{i+1}=\sum_{i=1}^{n-1}
e^{i(\phi_{i+1}- \phi_i)}.$$

As is usual with chiral Hamiltonians the meaning of the chiral action
is a bit imprecise.  To make it more precise we note that the usual
way is to consider insertions of $\int J_+$ on cyclinders
and by gluing rules extend it to higher genera.

It is convenient to also describe the deformations of the $A_{n-1}$
singularity in this formalism.  If we bosonize the fermion
$\psi_i= e^{i\phi_i}$ as before, the above argument shows that the
momentum $\partial \phi_i$ corresponds to the eigenvalue of the action
of the operator $x$ on the corresponding brane.  Here it is important
to consider the brane in the background of the screening operators
$\exp\int J_+$.  In this case the action of $x$ on the
$\psi_i$ is again strictly given by the $\partial \phi_i$ insertion.
Thus $x$ acting on the $i$th brane has eigenvalue $p_i$.  Therefore we
see that $x$ acting on the collection of branes satisfies
$$
\prod_{i=1}^n (x-p_i)=0,
$$
which describes the general deformation of the $A_{n-1}$ singularity.
We thus see that the Coulomb parameters of the $U(n)$ gauge theory get
naturally identified with the momenta of the corresponding scalar
fields.  In particular if we write the above equation as
$$
\sum_{k=0}^n x^{n-k} \sigma_k(p)=0,
$$
then
\eqn\symp{
\sigma_k(p)=\sum_{distinct\  i_r}p_{i_1} \cdots p_{i_k}
=\sum_ {distinct\  i_r}\partial \phi_{i_1}\cdots\partial \phi_{i_k}}

\subsec{Chopping the $A_{n-1}$ Geometries and D-brane probes}

\ifig\intersec{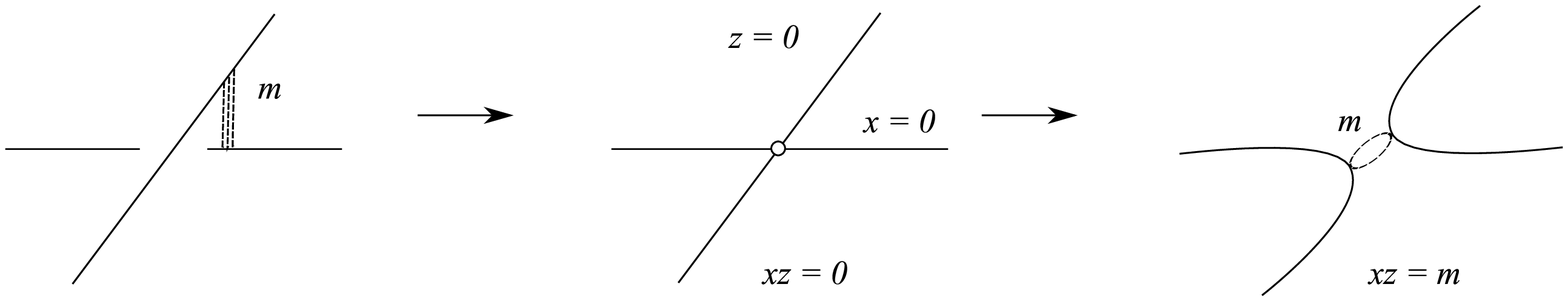}{120}
{Putting $m$ B-branes on the resolved conifold deforms the curve
$xz=0$ to $xz=m$.}

Let us now turn our attention to the inclusion of D-branes.  Let us
consider the $A_0$ case first.  This is given by the local geometry of
the conifold
$$
uv + xz=0.
$$
The corresponding curve is given by $F(x,z)=xz=0$.  This is depicted
in \intersec.  

Before the transition we put $m$ B-branes wrapping the $\P^1$ that
appears in the resolution of the conifold singularity. These branes
create after the transition a hypermultiplet of mass $m$. After the
transition the curve is changed into
$$
xz = m.
$$
We are interested in the probe brane living on the z-plane. From the
perspective of the probe we have $m$ units of flux at $z=0$ created by
the B-branes.  As mentioned before, the flux is measured by the
holomorphic 3-form, which in the local curve model is captured by the
field $\partial \phi(z)$ with expectation value $xdz$.  Thus we see
that the effect of $m$ B-branes is to lead to a flux
$$
\oint_{z=0} \del \phi=mg_s.
$$
In other words, this corresponds to the insertion of the operator
$e^{im\phi}$ at $z=0$.  More generally, if we considered the geometry
given by $F(x,z)=x(z-q)=0$ and placed $m$ B-branes ending at $z=q$ we
would have an insertion of $e^{im\phi(q)}$.

Now we consider the $A_{n-1}$ geometries, ``chopped off'' at some values
of $z$, in other words local geometries of the form
$$
F(x,z)=x^n(z-q)=0.
$$
One can ask what is the effect of the mass insertion on the
transition.  Locally this can also be described in the following way.
Let us first concentrate on the $n=2$ case.  If $m=0$, the geometry is
simply given by
$$
x^2(z-q)=0.
$$
However, the mass deformation will induce a transition in the geometry
near $z=q$.  It is not difficult to find out wat this effect should
be, namely it should be the mirror of the local blow up.  We claim it
is given by
$$
x^2(z-q)-mx=0.
$$
\ifig\sutwo{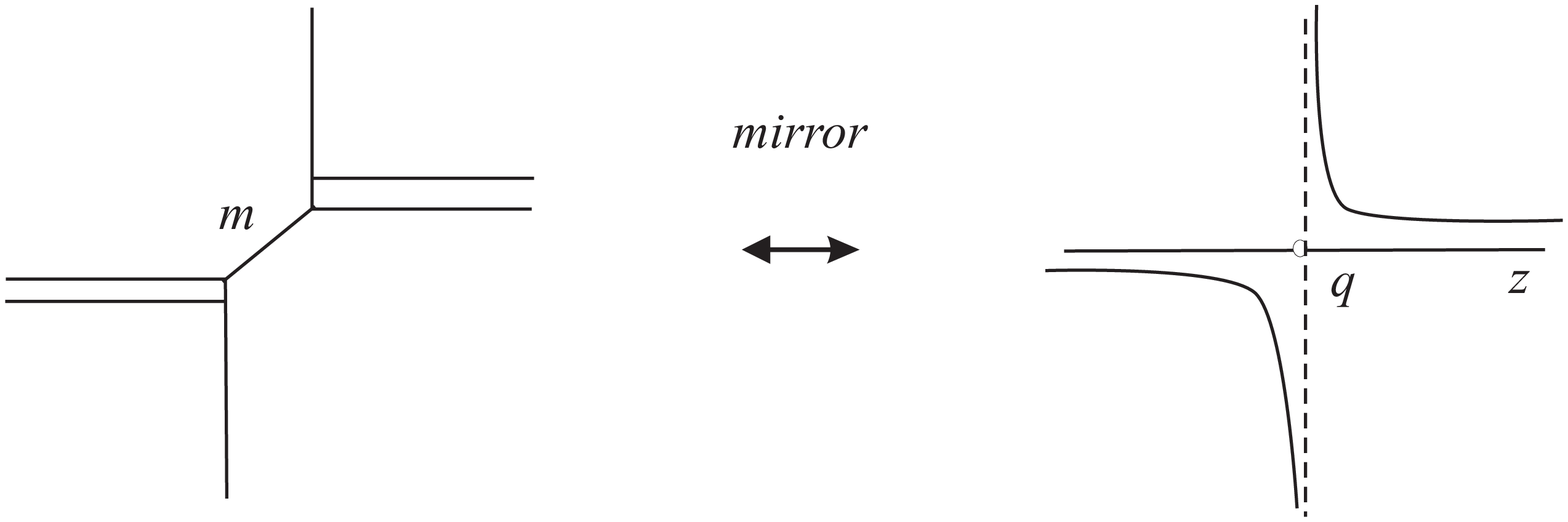}{100}
{Putting $m$ branes on the $A_1$ curve that describes a $SU(2)$ gauge
theory, gives the curve $x^2(z-q)-mx=0$.}

To see this, note that if we consider the curve $x^2(z-q)-mx=0$, and
study how $x$ varies as a function of $z$, we see that for large
values of $z$ the two roots are on top of each other.  However, near
$z=q$ the roots are split.  The mirror geometry is fixed uniquely by
having the correct asymptotics of the spectral curve (matching the
A-model toric geometry).  This is the geometry depicted in \sutwo.

Using the relation of momenta to the expectation values of
$\partial \phi_i$ given in \symp\ we see that this means
$\partial \phi_1+\partial \phi_2=m/(z-q)$, but
$\partial \phi_1 \cdot \partial \phi_2=0$.  This mass can be induced
by inserted the operator $\exp(im\phi_1(q))$. (The other choice can be
obtained from this by permutation action.)

\ifig\sun{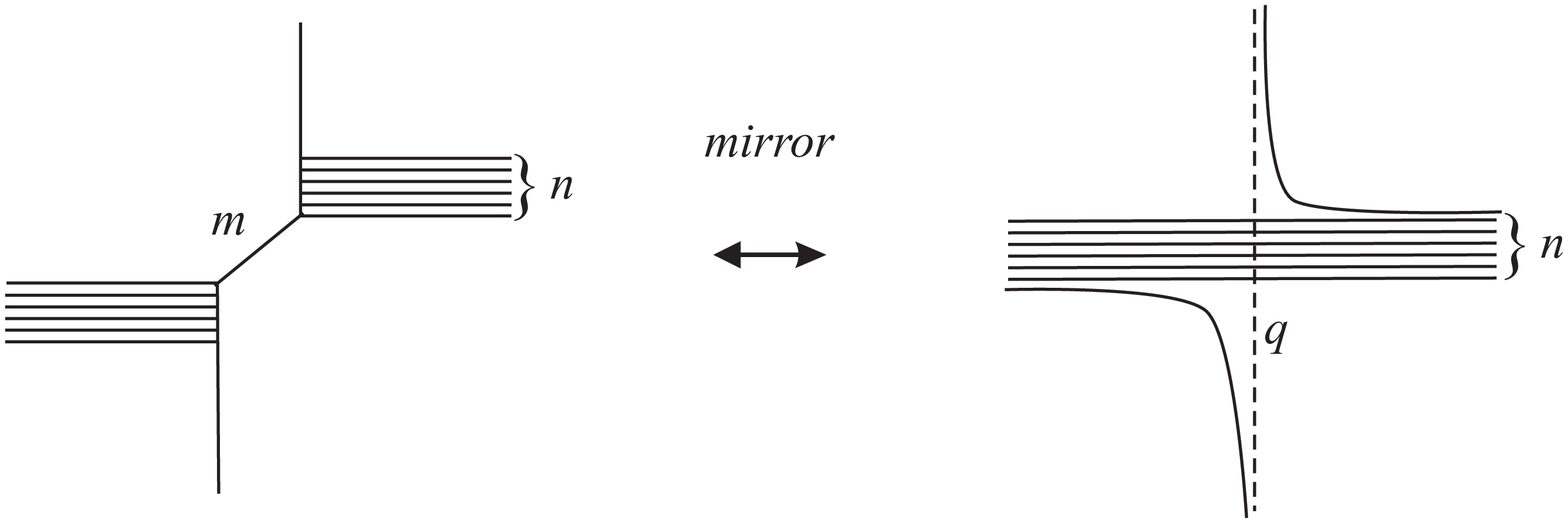}{100}
{The generalization of \sutwo\ to a $A_{n-1}$ curve that describes a
$SU(n)$ gauge theory. Now the curve is given by $x^n(z-q)-mx^{n-1}=0$.}

Similarly we can consider the more general $A_{n-1}$ case.  In that
case the B-model mirror for the mass insertion is uniquely fixed by
matching the A-model asymptotics (see \sun), and is given by
$$x^n-{m\over{z-q}}x^{n-1}=0$$
which is again induced by the insertion of $\exp(m \phi_1(q))$ in the
the path integral.  Note that this agrees with the prescription
suggested in \wyll.

It is sometimes convenient to shift variables, to write the curves in
a slightly different form.  For example, note that for the $A_1$ case
we can shift $x\rightarrow x- {1\over 2} {m\over z-q}$. Then the curve
can be written as
$$
x^2 = \left({m \over 2(z-q)}\right)^2.
$$

In a similar vein we see that in the presence of several masses $m_i$
at positions $q_i$ we effectively have a different CY geometry after
transition given by
$$
uv + x^2-x\sum_{i} {m_i\over z-q_i}=0.
$$
In the case of genus zero this is exact, {\it i.e.}, here the Green's
functions are simple and given by $1/(z-q_i)$.\foot{ It is easy to generalize
these statements for higher genus.}  By a change of variables
$x\rightarrow x-\sum_i {1\over 2} {m_i\over z-q_i}$ we end up with the
geometry
$$
uv + x^2-W'(z)^2=0,
$$
where
$$
W(z)=\sum_i {m_i\over 2} \log (z-q_i).
$$
We thus end up with a local Calabi-Yau which has conifold like
singularities at the points where $W'(z)=0$. 

However, this geometry is still not the most general geometry relevant
for the gauge theory in question, because we would be interested in
turning on arbitrary Coulomb branch parameters.  That will resolve the
conifold singularities.  These turn out to correspond to bringing down
the screening operators $\int J_+$.  This naturally connects to the
matrix model description discussed in section 2 where we found that
the insertion of $\int J_+$ is related to placing branes (matrix
eigenvalues) at the critical points of $W$.

We have thus receovered the equivalence of the Brane probe theory
with the Toda theory together with the correct dictionary between them.  Namely
we have a $U(n)$ Toda theory with mass insertions given by
fundamental weight of Toda.  This derivation was done for the case where
 the Nekrasov deformation is turned off.  In the next section
we extend this derivation to this deformation which leads to turning on
the background charge for Toda.

\newsec{Nekrasov's Deformation and $\beta$-Ensembles}

We will now generalize the above relations between topological
strings, matrix models and Toda theories to include Nekrasov's
deformation of the topological strings. From the point of the four
dimensional gauge theories this corresponds to working equivariantly
with respect to the $U(1) \times U(1)$ rotation group inside the
$SU(2) \times SU(2) \cong SO(4)$ acting on the space-time $\R^4$. The
corresponding equivariant parameters $\e_1,\e_2$ have a physical
interpretation as the angular velocities of these rotations. The
geometric implementation involves the so-called $\Omega$
background \nek.  It is known that this corresponds to turning on
$SU(2)_R$ holonomies on the internal Riemann surface $\Sigma$ in order
to preserve the topological supersymmetry, that otherwise would be
broken by turning on the $\e_i$'s. More precisely: Nekrasov's
deformation corresponds to a two-form in $\R^4$ with self-dual
component proportional to $\e_1 - \e_2$ and anti-self-dual component
proportional to $\e_1 + \e_2$. So, if $\e_1 \not = - \e_2$
supersymmetry is broken.

Before we turn to the implications for topological string theory and
matrix models, let us first introduce some convenient notation to
parametrize the Nekrasov deformation. We will use the following
notation:
$$
b^2 = {\e_1 \over \e_2} = - \beta,
$$
and 
$$
g_s^2 = -\e_1 \e_2. 
$$
The Liouville background charge is then expressed as
$$
Q = b + {1\over b} = \sqrt{\e_1 \over \e_2} + \sqrt{\e_2\over \e_1} =
{\e_1 + \e_2 \over g_s}.
$$
Note that the duality symmetry $\e_1 \leftrightarrow \e_2$ corresponds
to $\beta \leftrightarrow {1 / \beta}$ and $b \leftrightarrow {1/ b}$.
The undeformed case is obtained at $\epsilon_1 = - \epsilon_2 = g_s$,
$b=i$, $Q=0$.

\subsec{$\Omega$ backgrounds and Toda field background charges}

Let us first consider the implications for this deformation for the
geometry of topological strings. We interpret Nekrasov's $\Omega$
background in the present context as follows: Let us consider the
spacetime to be $\R^4={\bf C}^2$ with coordinates $z_1,z_2$.  Consider
the canonical line bundle of this complex spacetime represented by the
section $dz_1\wedge dz_2$.  The local geometry of Calabi-Yau over the
$\Sigma$ is
$$
{\cal N}_\Sigma \oplus {\bf C}^2.
$$
In other words, we view ${\bf C}^2$ also as part of the `internal'
geometry.  We are interested according to \nek\ in considering a
non-trivial bundle structure of ${\bf C}^2$ over $\Sigma$.  Let
$F_{{\bf C}^2}$ represent the curvature of the canonical line bundle
of ${\bf C}^2$ over $\Sigma$.  We take this to be represented by a
multiple of the curvature on $\Sigma$:
$$
F_{{\bf C}^2}=-\epsilon\cdot R_{\Sigma},
$$
where we will identify $\epsilon=\epsilon_1+\epsilon_2$.  Now,
however, the Calabi-Yau condition is lost and we have lost
supersymmetry.  To restore it, we need the ${\cal N}_\Sigma$ to also
twist. The trace of the rotation generator in ${\cal N}_\Sigma$ we
will denote by $U(1)_R\subset SU(2)_R$.  We need the curvature of this
bundle to cancel this additional curvature.  In other words we need to
change this curvature so that
$$
\delta F_{U(1)_R} = \epsilon\cdot R_{\Sigma}.
$$
Let $J_R$ denote the $U(1)_R$ current acting on fields living on
$\Sigma$.  This means that we need to add to the action of these
fields an additional term given by
$$(\epsilon_1+\epsilon_2)\int_\Sigma J_R \wedge \omega_\Sigma$$
where $\omega_\Sigma$ represents the spin connection on $\Sigma$.

To implement this deformation we need to identify $J_R$ in our
context.  Note that the $U(1)_R$ can be identified with the phase of
the holomorphic 2-form $dx \wedge du/u$ on the geometry normal to $\Sigma$.
This corresponds to action
$$
x\ \to\  \exp(i\theta)\cdot x,
$$
where the phase $\theta$ is given by
$$
\theta ={\epsilon_1+\epsilon_2 \over g_s} ={\epsilon_1+\epsilon_2 \over
\sqrt{\epsilon_1\epsilon_2}}.
$$
Such an action corresponds to the following transformation on the
fermions
$$
\psi_k \ \to \exp\left(i(k-1+const)\theta\right) \cdot \psi_k,
$$
using the fact that the $\psi_k$ are obtained from $\psi_1$ by the
action of $x^{k-1}$. The constant here depends on the choice of the
action of $U(1)_R$ on $\psi_1$.  A natural choice for this constant,
suggested by CTP invariance of the analog LG models, is $-(n-1)/2$.
In other words, this is like twisting the LG model and therefore leads
to the background charge coupling given by
$$
S = \ldots + i \theta \int \sum _k Q_k \partial \phi_k \wedge \omega_\Sigma 
= \ldots + i\theta \int \sum _k Q_k  \phi_k  R^{(2)} \sqrt{g},
$$
where $Q_k=(k-(n+1)/2)$ and 

This implies that we have a term proportional to
$$
\int (b+1/b) \sum_k (k-(n+1)/2) \phi_k R^{(2)} \sqrt{g}, 
$$
where $ b=\sqrt{\epsilon_1/\epsilon_2}$. We thus see we have obtained
the Toda system with background charge, simply by considering twisting
of the normal geometry over $\Sigma$.

\subsec{Generalized matrix models and $\beta$ ensembles}

We will now turn to the corresponding deformation in the matrix model,
which will turn out to be given by the so-called
$\beta$-ensemble. Again, as in section 2, for simplicity we restrict to
the $SU(2)$ case, but generalizations to $SU(n)$ are straightforward.

As we have just seen, including general $\e_1,\e_2$ corresponds to a
background charge $Q$ for the scalar field $\phi$. This changes the
conformal dimensions of the vertex operators. In particular, the
screening charge (of dimension 1) is now given by
$$
J_+ = e^{b\phi}.
$$
Correlations of the Coulomb gas are now given by
$$
\bigl\langle J_+(z_1) \cdots J_+(z_N) \bigr\rangle = \prod_{I<J} (z_I
- z_J)^{-2b^2} = \Delta(z)^{-2b^2}.
$$
Replacing the usual second power of the Vandermonde determinant by
this measure is known in the matrix model literature as the
$\beta$-ensemble \refs{\mehta,\bet}. 
These generalized matrix models are given by
eigenvalue integrals of the form
$$
Z = \int d^N\!z \, 
\Delta(z)^{2\beta} \cdot  \exp \sum_I {\beta \over g_s} W(z_I).
$$

Clearly, we can map the Liouville model directly to the
$\beta$-ensemble if we identify
$$
\beta = -b^2.
$$
These generalized measures are traditionally studied because they
naturally appear if, instead of a unitary ensemble of hermitean
matrices, one considers orthogonal or symplectic matrices. More
precisely, the $\b$-ensemble reproduces these alternative marix models
for
$$
SO(N):\ \beta={1\over 2}, \qquad
Sp(N):\ \beta=2.
$$
Note that for these values the duality symmetry $\beta \to 1/\beta$
interchanges
$$
SO(N) \leftrightarrow Sp(N),
$$
which is not an unfamiliar symmetry for string theory.

We claim that for any geometry on which the topological string can be
described by a matrix model, generalizing the measure to the
$\beta$-ensemble immediately gives the refined invariant as a function
of $\e_1,\e_2$. This becomes already clear by considering the case of
a Gaussian measure which corresponds to the conifold geometry. We find
$$
Z =  \int d^N\!z \ \Delta(z)^{2\beta}  \cdot \prod_I e^{-z_I^2}
$$
This integral is a special case of the Selberg integral and can be
evaluated to be
$$
Z = (2\pi)^{N/2} \prod_{k=1}^N {\Gamma(1 + \beta k) \over \Gamma (1 + \beta)}.
$$
This gives for the free energy in the large $N$ limit
$$
F = \log Z = \int {ds \over s} {e^{\mu s} \over (1 - e^{\e_1s})(1 -
e^{\e_2s})}
$$
with $\mu=g_sN$. This is indeed the contribution of a single
hypermultiplet of mass $\mu$.  Note that we recognize in this expression
the partition function of the $c=1$ string at radius $b$. This
therefore generalizes the usual identification of the conifold with
the $c=1$ string at self-dual radius for $\beta = -b^2 = 1$. The
symmetry $b \to 1/b$ is also obvious.

To a large extent the methods that are used to solve traditional
matrix models can be generalized to the $\beta$-ensemble. For example,
the loop equations again are a reflection of diffeomorphisms $\delta z
= z^{n+1}$ acting on the eigenvalues as generated by the modes $L_n$
of the stress-tensor $T(z)$. However, because of the `funny'
$\b$-measure, the stress-tensor is no longer quadratic in terms of the
collective field
$$
\d\phi = W'(z) + \sum_I {g_s \over z_I -z}.
$$
It picks up a background charge \bet
$$
T(z) = \half (\d\phi)^2 + Q \d^2\phi,
$$
with $Q=b+1/b$. This is of course perfectly consistent with the
connection to Liouville theory that we have argued for. It can be seen
as a confirmation that we are implementing the
Nekrasov deformation correctly in the topological string framework.

So, combining all ingredients we see that the $\b$-ensemble can be
mapped to a chiral version of Liouville with general central charge
$c=1+ 6Q^ 2$
$$
\int d^2\!z \left(\d\phi\dbar\phi + Q \phi R^{(2)} \sqrt{g} 
\right) + \int dz \, e^{b\phi}.
$$

The presence of a background charge does not influence the leading
solution of the matrix model, which is captured by the quadratic part
of $T(z)$. The genus zero contribution $\cF_0$ is therefore still
described in terms of the spectral curve \hyperell. However, this is
not rue for contributions with a different topology. Because $T(z)$ is
no longer invariant under $\phi \to -\phi$, the partition function as
a function of $g_s$ now also receives odd contributions
$$
Z = \exp F, \qquad F = \sum_{n \geq 0} g_s^{n-2} \cF_{n \over 2}(\beta).
$$
This is well-known for the $SO/Sp$ matrix models at $\b={1\over 2},2$
where unoriented surface contribute. For example $\cF_{1 \over 2}$
represents the contribution of $\R\P^1$.

This connection can be extended to correlation functions of vertex
operators. They take the form
$$
V_m(q) = e^{m\phi/2} = \prod_I (z_I -q)^{-bm}
$$
Thus we have to analyse the expression
$$
\Bigl\langle \prod_i V_m(q_i) \Bigr\rangle_{\strut N}
= \int d^N\!z\, \prod_{I < J }\left(z_I
- z_J\right)^{-2b^2}
\prod_{i, I} \left(z_I - q_i\right)^{-bm_i} 
$$
in the limit $N,m_i \to \infty$. The analysis of this model follows
exactly along the same lines as in section 2 for $\b=1$.

\newsec{Directions for Future Research}

In this paper we have shown that the AGT conjecture about the chiral
blocks of Toda CFT being related to $\cN=2$ gauge theories, arises
naturally from its interpretation in the context of topological
strings.  A key ingredient in this derivation is the large $N$ limit
of branes in topological strings and the geometric transitions they
induce.  We gave two (not completely unrelated) derivations: One was
more conceptual, based mainly on the general properties of B-branes
and their relation to free fermionic systems and non-commutative
geometry. The other was more specific to the relation between
topological strings and matrix models.  In particular we related
this to the matrix models of `multi-Penner' potentials for the case of sphere.

Let us now mention some points for further study. First of all, it is
important that the SW geometry arises for us holographically in the
large $N$ limit of the brane probe theory, even for $A_1$.  There is
yet another sense of large $N$ limit one can consider in this context,
which is the $A_{n-1}$ geometry in the limit of large $n$.  The
gravity version of this has been studied in \gaiomalda.  Combining
this with our approach we have a double $N$ holography.  In the
context of matrix models, this corresponds to the large $N$ limit of
$A_{\infty}$ quiver matrix models.  As discussed in \dvs\ this
corresponds to matrix quantum mechanics.  It would be interesting to
develop this limit of the theory from this viewpoint.

Second, it is clear that the $D$ and $E$ series can be studied along
the lines suggested here.  In particular the corresponding matrix
models would be a quiver of the $D$ or $E$ type.  It should be
interesting to develop techniques to better understand these matrix
models, or equivalently the corresponding Toda correlations.

Third, we have interpreted Nekrasov's $\Omega$ background in a
specific geometric way in topological strings based on a local
curve. It would be intersting to study this further, and in particular
consider its application to more general local Calabi-Yau geometries.

Finally, in this paper we have focused on connections between
topological strings and 4d theories.  However, it is well known that
topological strings also captures vacuum geometry of 5d gauge theories
compactified on a circle, more clearly realized in the context of type
IIA on CY 3-folds in terms of toric geometry. In a sense here we have
been studying only the special limit $R\rightarrow 0$ of such a system.
Clearly it is important to study the $R$ dependence of this bigger
system.  The most naive generalization of our results would suggest,
in the matrix model setup, to replace the Vandermonde by a
$q$-deformed version.  It would be interesting to study these issues
further.

\vglue 1cm

\noindent{\bf Acknowledgements}
\smallskip

We would like to thank Davide Gaiotto, Sergei Gukov, Martin Rocek, Piotr
Sulkowski, and Herman Verlinde for valuable discussions.

This research was initiated during the Seventh Simons Worshop on
Mathematics and Physics.  We thank the Simons Center for Geometry and
Physics for providing a stimulating research environment as well as
for its warm hospitality. The research of R.D. was supported by a NWO
Spinoza grant and the FOM program {\it String Theory and Quantum
Gravity}. The research of C.V. was supported in part by NSF grant
PHY-0244821.

\listrefs

\end